\documentclass{jfm}
\usepackage{graphicx}
\usepackage{epstopdf, epsfig}
\usepackage{amsmath,amssymb,amsfonts,bm}
\usepackage{rotating}
\usepackage{color}
\usepackage{mathtools}
\usepackage{subfig,setspace}

\newcommand{\pd}[1]{\partial_{#1}}
\newcommand{\ov}[1]{\overline{#1}}

\newcommand{\eps}{\epsilon}

\newcommand{\Ro}{Ro_f} 
\newcommand{\Eu}{Eu_f} 
\newcommand{\Fr}{Fr_f} 

\newcommand{\Pra}{\sigma} 

\newcommand{\wt}[1]{\widetilde{#1}}
\newcommand{\mbf}[1]{\boldsymbol{#1}}
\newcommand{\mcl}[1]{\mathcal{#1}}

\newcommand{\ra}{\rangle}
\newcommand{\la}{\langle}
\newcommand{\veps}{\varepsilon}

\shorttitle{Non-Hydrostatic, stably-stratified and rapidly rotating flows}
\shortauthor{D. Nieves, I. Grooms, K. Julien,  and J. Weiss}

\title{Investigations of non-hydrostatic, stably stratified and rapidly rotating flows}

\author{David Nieves\aff{1}
\corresp{\email{david.nieves@colorado.edu}}
  \and Ian Grooms\aff{1} \and Keith Julien\aff{1} \and Jeffrey B. Weiss\aff{2}}

\affiliation{\aff{1}Department of Applied Mathematics, University of Colorado,
Boulder, CO 80309, USA \\ 
\aff{2}Department of Atmospheric and Oceanic Sciences, University of Colorado,
Boulder, CO 80309, USA}

\begin{document}

\maketitle

\begin{abstract}
We present an investigation of rapidly rotating (small Rossby number $Ro\ll 1$) stratified turbulence where the stratification strength is varied from weak (large Froude number $Fr\gg1$) to strong ($Fr\ll1$). 
The investigation is set in the context of a reduced model derived from the Boussinesq equations that efficiently retains anisotropic inertia-gravity waves with order-one frequencies 
and highlights a regime of wave-eddy interactions. Numerical simulations of the reduced model are performed where energy is injected by a stochastic forcing of vertical velocity, which forces wave modes only.
The simulations reveal two regimes characterized by the presence of well-formed, persistent and thin turbulent layers of locally-weakened stratification at small Froude numbers, and by the absence of layers at large Froude numbers. Both regimes are characterized by a large-scale barotropic dipole enclosed by small-scale turbulence. When the Reynolds
number is not too large a direct cascade of barotropic kinetic energy is observed, leading to total energy equilibration. We examine net energy exchanges that occur through vortex stretching and vertical buoyancy flux
and diagnose the horizontal scales active in these exchanges. We find that the baroclinic motions inject energy directly to the largest scales of the barotropic mode,
implying that the large-scale barotropic dipole is not the end result of an inverse cascade within the barotropic mode.
\end{abstract}

\begin{keywords}
Authors should not enter keywords on the manuscript, as these must be chosen by the author during the online submission process and will then be added during the typesetting process (see http://journals.cambridge.org/data/\linebreak[3]relatedlink/jfm-\linebreak[3]keywords.pdf for the full list)
\end{keywords}

\section{Introduction}
The study of fluid turbulence connects bulk statistical properties like energy spectra, structure functions, and the energy dissipation rate to physical processes like vortex stretching and instabilities \citep{Frisch95}.
In the context of geophysical turbulence, the emphasis is on how rotation and density stratification affect the statistical and dynamical properties of the turbulent flow.
At small scales rotation and buoyancy are expected to become dynamically unimportant, with statistics resembling those of non-rotating, constant-density flow.
More specifically, rotation and buoyancy respectively are expected to become unimportant for scales smaller than the Zeman scale  $L_\Omega = \sqrt{\epsilon/(2\Omega)^3}$ \citep{Zeman94} and the Ozmidov scale $L_N = \sqrt{\epsilon/N^3}$ \citep{Ozmidov65} where $\epsilon$ is the mean rate of energy dissipation per unit mass, $\Omega$ is the rate of rotation, and $N = \sqrt{-g\partial_z\rho/\rho_0}$ is the buoyancy frequency in a density-stratified fluid under the Boussinesq approximation ($g$ is the gravitational acceleration, $\rho$ is the density, and $\rho_0$ is a constant reference density). 
Studies of geophysical turbulence therefore include scales larger than either the Ozmidov or Zeman scales, or both.

Rotation and stratification induce restoring forces that lead to wave dynamics; when the axis of rotation is parallel to gravity, the linear wave spectrum includes frequencies between $N$ and $2\Omega$.
Rotation and stratification are expected to have a qualitative impact on turbulence when the period of wave dynamics is comparable to or less than the time scale of nonlinear advection.
More precisely, rotation and stratification respectively are expected to strongly affect the dynamics when the Rossby number Ro $=U/(2\Omega L)$ and Froude number Fr $=U/(NL)$ are small, where $U$ and $L$ are characteristic velocity and length scales of the turbulent flow.
Geophysical turbulence is characterized by small Rossby and/or Froude numbers.

The linear eigenfunctions of the Boussinesq system include two wave modes and a zero-frequency `vortical' mode \citep{Bartello95}.
At low Rossby and Froude numbers there is a clear time scale separation between the slow, nonlinear evolution of the vortical mode and the fast, weakly-nonlinear evolution of the wave modes which can be exploited to derive asymptotically a reduced set of dynamics for the vortical modes; this reduced system is the celebrated quasigeostrophic equations \citep{Eady_1949,Charney_1948,Pedlosky_87,Vallis06}.
Time scale separation was exploited by Embid and Majda to rigorously prove the validity of the quasigeostrophic system even in the presence of wave modes with amplitudes comparable to the vortical modes, in contrast to the asymptotic derivation which assumes that any waves have low amplitude \citep{EM_CPDE_1996,EM_GAFD_1998,ME_TCFD_1998}.
\citet{TW_SJMA_2010,TW_JAS_2011} have also proven rigorously that, under mild assumptions, the small-Rossby, small-Froude dynamics eventually approaches a quasigeostrophic balance irrespective of the amplitude of wave modes in the initial condition.
The quasigeostrophic system is thus a natural touchstone for geophysical turbulence, and the qualitative properties of turbulence in the quasigeostrophic system were presciently forecast by \citet{C_JAS_1971} based on an analogy with previous studies of two-dimensional turbulence.

The rigorous framework of \citet{EM_CPDE_1996} exploits an asymptotic time scale separation between the fast wave dynamics and the slow `balanced' dynamics.
\citet{EM_GAFD_1998} and \citet{WEHT_JFM_2011} also used the framework to rigorously derive equations governing the slow limiting dynamics in the limits of low Froude and finite Rossby numbers, and low Rossby and finite Froude numbers, respectively.
Because of the need for an asymptotic time scale separation, the slow limiting dynamics include a single pair of wave modes at the slowest linear frequency ($2\Omega$ for \citet{EM_GAFD_1998} and $N$ for  \citet{WEHT_JFM_2011}) and all other wave modes are assumed to be asymptotically fast by comparison, and do not appear in the slow limiting dynamics.
Results analogous to those of \citet{TW_SJMA_2010} for the quasigeostrophic system are lacking for these two systems of slow limiting dynamics, and it is not yet clear whether these systems have the same relevance for geophysical turbulence in their respective asymptotic regimes as the quasigeostrophic system has for the low-Froude, low-Rossby number regime.

If either the Rossby or Froude number is order-one, there is not a clear time scale separation between the linear wave dynamics and the nonlinear advective dynamics, so a reduced system that eliminates nearly all the wave dynamics is arguably inappropriate.
Nevertheless, the smallness of one of the nondimensional numbers can still be exploited in both cases to reduce the complexity of the full Boussinesq system.
When the Froude number is small but the Rossby number is order-one one can make the hydrostatic approximation to arrive at the so-called primitive equations.
When the Rossby number is small but the Froude number is order-one one can make a geostrophic approximation and arrive at the non-hydrostatic quasigeostrophic equations \citep[NHQGE;][]{JKW_TCFD_1998,JKMW_JFM_2006}.
Both of these equation sets are significantly easier for both numerical simulation and mathematical analysis than the unreduced Boussinesq equations, and both sets of equations include linear wave dynamics with frequencies between either $2\Omega$ (primitive equations) or $N$ (NHQGE) and infinity.
The quasigeostrophic equations can be recovered from both sets of equations in the limit where both the Rossby and Froude numbers are small.

As the Froude number is typically smaller than the Rossby number in atmospheric and oceanic turbulence, studies of rotating, stratified turbulence have primarily focused on strongly-stratified regimes where the Froude number is small.
The regime of geostrophic turbulence with low-Rossby number and order-one Froude number has seen comparatively little study, though this regime is relevant to weakly stratified abyssal ocean dynamics at high latitudes and in the western Mediterranean \citep{ELM_JPO_1984,TGC_DSR_2003,HM_GRL_2005,TMR_JPO_2007}.
The regime is also relevant to planetary and stellar interiors where the stratification transitions from unstable (imaginary $N$) to stable ($N\ge0$).
Examples include the solar tachocline believed to be the origin of large scale solar magnetism \citep{mM05}
and the Earth's outer liquid core where the existence of stably-stratified layers have been postulated
\citep{pozzo2012thermal}.
The present investigation focuses on rotating, stratified turbulence at low Rossby number, with Froude numbers varying from large to small.

The main points of comparison for the transitional regime of Froude numbers between zero and infinite are the `quasigeostrophic' regime at small Froude numbers and pure rotation at large Froude numbers.
Quasigeostrophic turbulence theory, by analogy with the theory of two-dimensional turbulence \citep{BE_ARFM_12}, predicts a transfer of energy from the forcing scale to larger scales through an inertial range where the energy spectrum is proportional to $\tilde{k}^{-5/3}$, where $\tilde{k}^2=k_h^2+(2\Omega/N)^2k_z^2$.
At scales smaller than the energy forcing quasigeostrophic turbulence theory predicts an energy spectrum proportional to $\tilde{k}^{-3}$.
The $-5/3$ \citep{SW_JFM_2002,MMRP_EPL_2013} and $-3$ \citep{WB_JFM_2006} spectral slopes are evident in simulations of triply periodic Boussinesq dynamics in the regime of low Rossby and Froude numbers, and both \citet{WB_JFM_2006} and \citet{WW_JFM_2014} observed energy accumulating in the vortical modes.
These results underscore the importance of quasigeostrophic dynamics, and demonstrate that the theorem of \citet{TW_SJMA_2010} applies qualitatively even in this stochastically-forced regime.

In simulations of constant-density (infinite Froude number) low-Rossby number turbulence energy is transferred to scales larger than the forcing scale through an inertial range with spectrum proportional to $k^{-3}$; energy is also primarily transferred to a depth-independent horizontal velocity, the `barotropic mode' \citep{SW_PoF_1999,SL_JFM_2005}.
\citet{MMRP_EPL_2013} found transfer of energy into the barotropic mode to be less rapid in the purely rotating regime than in the quasigeostrophic regime.
\citet{SMRP_PRE_2012} observed a $k^{-5/3}$ spectrum at large scales in a purely rotating system when the stochastic forcing was depth-independent; this case is somewhat degenerate and likely not indicative of universal behavior.

The transitional regime between pure rotation and quasigeostrophy has seen comparatively few simulations.
In the experiments of \citet{SS_GAFD_2008} the Froude and Rossby numbers are both comparatively small, though in some experiments the Froude number was larger by up to a factor of 5.
In their experiments with Froude number larger than Rossby number the wave mode energy grows to dominate, in contrast to the behavior in both the quasigeostrophic and purely-rotating limits where energy accumulates primarily in the vortical and depth-independent components; this may be related to the fact that forcing was applied near the scale of the computational domain.
They found that the vortical mode spectrum retained its quasigeostrophic $k^{-3}$ behavior at scales smaller than the forcing, though it deviated towards a shallower slope at much smaller scales.
In a single experiment with low Rossby number and moderate Froude number, also forced near the scale of the computational domain, \citet{AK_EPL_2011} diagnosed a downscale transfer of both energy and potential enstrophy; spectral slopes were not reported.
\citet{WW_JFM_2014}  also forced near the scale of the computational box, and found energy accumulating in the barotropic mode; spectral slopes were not reported.
These investigations leave open entirely the question of how the large scale dynamics transition between the quasigeostrophic and purely rotating regimes as the Froude number increases, which is the focus of the present investivation.

The paper is organized as follows: \S~\ref{sec:govn_eqns} introduces preliminaries including discussions regarding Proudman-Taylor constraints and inertia-gravity waves, \S~\ref{sec:deriv_redeqns} provides an overview of the reduced equations used in our numerical simulations, \S~\ref{sec:num_meth} summarizes the numerical methods including the forcing scheme employed for numerical simulations, and \S~\ref{sec:results} gives the results of our numerical experiments.

 
\section{Governing Equations and Preliminaries}\label{sec:govn_eqns}

We consider an incompressible fluid subject to an imposed constant vertical gravitational field $\mbf{g} = -g\hat{\mbf{z}}$ 
and a system rotation with constant angular velocity $\mbf{\Omega} = \Omega\hat{\mbf{z}}$. The fluid is stably-stratified in the vertical 
with total density $\rho^{*}=\hat{\rho}^{*}(z^*)+\rho'^{*}(\mbf{x}^{*},t^{*})$, where $\hat{\rho}^{*}(z^*)=\rho^{*}_{0} +\delta\hat{\rho}^{*}(z^*)$ is an ambient density profile consisting of a
constant reference density $\rho^{*}_{0}$ and a density variation $\delta\hat{\rho}^{*}(z^*)$ (where asterisks denote dimensional quantities). It follows that the total buoyancy of a fluid parcel, given by
\begin{equation}
{\mathrm{b}}^{*} = -\frac{g}{\rho_0^{*}}\left(\delta\hat{\rho}^{*}(z^*)+\rho'^{*}(\mbf{x}^{*},t^{*})\right) = -\frac{g}{\rho^{*}_{0}}\delta\hat{\rho}^{*}(z^*) + b^{*\prime}(\mbf{x}^{*},t^{*}),
\label{eqn:buoy_decomp}
\end{equation}
\noindent is decomposed as the sum of the ambient buoyancy field and a fluctuating component $b^{*\prime}$
associated with fluid motions. Pressure is decomposed in a fashion similar to buoyancy $p^{*} = \hat{p}^{*}(z^{*})+p'^{*}(\bm{x}^{*},t^{*})$ with a pressure component 
in hydrostatic balance with the ambient buoyancy
\begin{equation}
\partial_{z^*} \delta \hat{p}^{*}(z^{*}) = - g \hat{\rho}^{*}(z^{*}).
\end{equation}
\noindent The governing equations in the Boussinesq approximation for a fluid with constant kinematic viscosity $\nu$ and  buoyancy diffusion $\kappa$ are given by 
\begin{subeqnarray} 
D^*_{t}\mbf{u}^* + 2 \Omega{\hat{\mbf{z}}} \times\mbf{u}^* &=& -\nabla p^{*\prime} +  b^{*\prime} \hat{\mbf{z}} + \nu \nabla^{*2} \mbf{u}^*, \\ 
D^*_{t} b^{*\prime}  + N^2(z^*) w^*  &=& \kappa \nabla^{*2}b^{*\prime} , \\ 
\nabla^* \cdot \mbf{u}^* &=& 0. 
\label{eqn:DBOUS_b}
\end{subeqnarray}
\noindent where
\begin{equation}
D^*_{t}(\bcdot) = \left[\partial_{t^{*}}+\mbf{u}^{*}\cdot\nabla^{*}\right](\bcdot).
\end{equation}
The ambient stratification is now characterized by the buoyancy (Brunt-V\"ais\"al\"a) frequency 
$N^{2}(z^*)=-g\rho^{*-1}_{0}\partial_{z^*}(\delta\hat{\rho}^{*}(z^*))$.

An external forcing is required to excite fluid motions, and in the present investigation energy is generated by a stochastic vertical velocity forcing.
Recent studies in a similar parameter regime have used stochastic buoyancy forcing \citep{WW_JFM_2014} or simultaneous forcing of all components of velocity \citep{MMRP_EPL_2013}.
The present investigation includes regimes of weak stratification (large Froude numbers), and initial tests with buoyancy forcing in the weakly-stratified regime led to frequent large-scale overturning.
Vertical velocity forcing avoids these spurious dynamics in the weakly-stratified regime while also avoiding direct forcing of the slow quasigeostrophic dynamics in the strongly-stratified regime.

Characteristic scales determined from the energy injection rate $\eps^{*}_{f}$ and forcing length scale $L^*_{f}$ are the forcing velocity, time, and buoyancy scales 
\begin{equation}
U^*_f = \left(\epsilon^{*}_{f} L^*_f\right)^{1/3}, \quad  T^*_f = \left(L_f^{*2}\eps_{f}^{*-1}\right)^{1/3}, \quad
B^*_f = \left(\eps^{*2}_{f}L_f^{*-1}\right)^{1/3}.
\end{equation}
\noindent This gives rise to the following nondimensional equations
\begin{subeqnarray} 
D_{t}\mbf{u} + \frac{1}{\Ro}\hat{\mbf{z}}\times\mbf{u} &=& -Eu_f\nabla p +  b \hat{\mbf{z}} + \frac{1}{Re_f} \nabla^{2} \mbf{u}, \\ 
D_{t} b + \frac{1}{\Fr^2} S\left  ( z\right  ) w &=& \frac{1}{\Pra Re_f}\nabla^{2}b \\ 
\nabla \cdot \mbf{u} &=& 0,
\label{eqn:BOUS_b_rescaled}
\end{subeqnarray}
where 
\begin{equation}
D_{t}(\bcdot) = \left[\partial_{t}+\mbf{u}\cdot\nabla\right](\bcdot)
\end{equation}
\noindent and $S(z)$ is the nondimensional stratification profile defined according to the relation
$N^{2} (z^*) = N^2_0 S( z)$. We have defined $N_0 \equiv |g\rho^{*-1}_{0}(\partial_{z^*} \delta\hat{\rho}^{*}(z^*))_{max}|$
as the maximal buoyancy frequency and $S \equiv -\pd{z}\delta\hat{\rho}$. 

The nondimensional parameters that appear in (\ref{eqn:BOUS_b_rescaled}) are determined a priori
based on the energy injection rate $\eps^{*}_{f}$ and forcing length scale $L^{*}_{f}$.
These parameters are the Rossby number $\Ro$, Froude number $\Fr$, Euler number $\Eu$, and Reynolds number $Re_f$ defined as
\begin{equation}
\Ro = \frac{U^*_f}{2 \Omega L^*_f}, \quad \Fr = \frac{U^*_f}{N^*_0 L^*_f}, \quad Eu_f = \frac{\delta p_0}{\rho_0 U_f^{*2}}, \quad
Re_f = \frac{U^*_f L^*_f }{\nu}\equiv\left ( \frac{\eps^{*}_f L^{*4}_f}{\nu^3}\right )^{1/3}.
\end{equation}
\noindent 
The Rossby number is the ratio of rotation period, $T^{*}_{\Omega} = 1/2\Omega$, to the forcing time, $T^{*}_{f} = 
L^*_f/U^*_f$, and measures the rotational constraint of the fluid at the forcing scale. Hereafter, we focus solely on the rotationally constrained  regime $Ro_f\ll1$. The Froude number is the
ratio of the Brunt-V\"ais\"al\"a  time, $T^{*}_{N} = 1/N_0$, to $T^{*}_{f}$ and measures the ratio of the slowest linear wave period to the nonlinear advective time scale.
The Reynolds number provides a nondimensional measure of the energy injection rate into the system
and therefore controls the degree of turbulence achieved at the forcing scale $L^{*}_{f}$.
The Euler number measures the significance of the pressure gradient force relative to inertial accelerations.
The Prandtl number $\sigma=\nu/\kappa$ is the ratio of dissipation parameters and quantifies the thermometric properties of the working fluid. 

In addition to the nondimensional forcing length scale $L_f=1$ four internal length scales are also present: the
dissipation (Kolmogorov) scale $L_{K}$, first Rossby radius of deformation $L_{D}$, the Zeman length scale $L_{\Omega}$, and the
Ozimodov length scale $L_{N}$. These
\textit{nondimensional} length scales are defined, respectively, as
\begin{eqnarray}
&L_{K}\equiv Re_f^{-3/4},\quad\hspace{7.5em}
&L_D \equiv  \left ( \displaystyle{\frac{N_0 H^*}{2 \Omega L^*_f}}\right ) =  \displaystyle{ \frac{A Ro_f}{\Fr}},\\
&L_{\Omega}\equiv \displaystyle{\left ( \frac{\eps^*_f L^{*-2}_f}{(2\Omega)^{3}}\right  )^{1/2} }= Ro_f^{3/2}\ll1,\quad
&L_N\equiv \displaystyle{\left ( \frac{\eps^*_f L^{*-2}_f}{N_0^{3}}\right)^{1/2}}= \Fr^{3/2}.
\end{eqnarray}
\noindent The dissipation scale is the scale at which the nonlinear turnover time equals the time scale of viscous dissipation.
The first Rossby radius of deformation is the scale where baroclinic instability converts potential to kinetic energy, and depends on $H^{*}$, 
the depth of the domain. The ratio $A=H^{*}/L_f^{*}$ is the nondimensional height of the domain.
In quasigeostrophic dynamics the conversion of baroclinic to barotropic\footnote{We adopt the convention that the `barotropic' component of the system includes only the depth-independent part of the horizontal velocity; all other fields including vertical velocity and buoyancy are `baroclinic.'} energy occurs mainly at scales larger than $L_D$. 
Rotation influences the dynamics at scales larger than the Zeman scale, and the Ozimodov scale is that above which eddies are influenced by
stratification.

In this paper, we consider only the case  $L_{\Omega} < L_{K}$ such that  all fluid scales are influenced by rotation.  This constraint places an upper bound for the Rossby number, namely
\begin{equation}
Ro_f = o\left (Re_f^{-1/2}\right ).
\end{equation}
Given that the Rossby number is very small and the Euler number passively scales pressure, it becomes clear that there exist two primary control parameters $Re_f, \Fr$. Varying these parameters causes the three dynamical length scales, $L_D$, $L_N$, and $L_K$ to vary through seven distinct regimes shown in Figure 1.
\begin{figure}
	\begin{center}
		\includegraphics[width=0.75\linewidth]{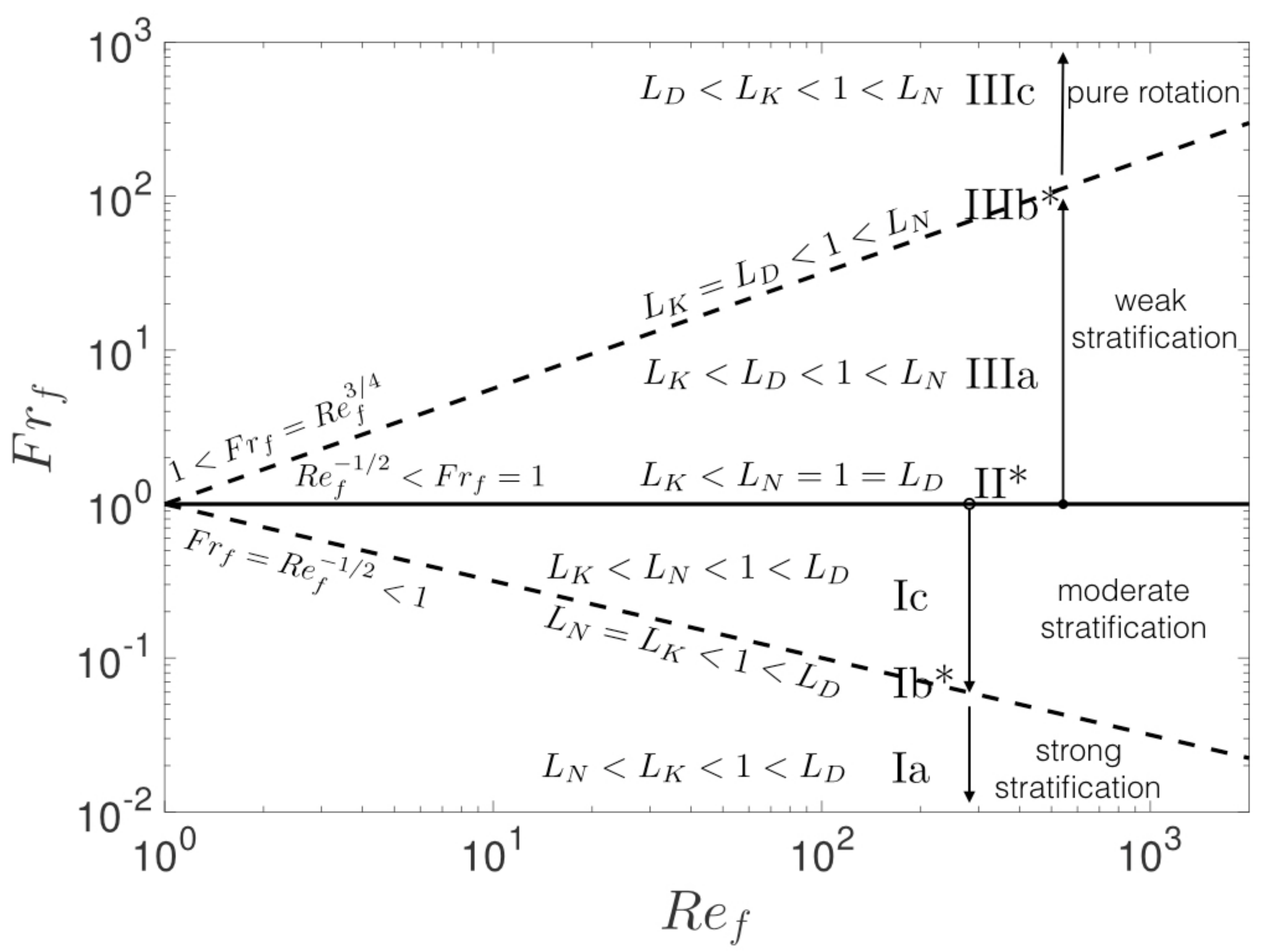} 
		\caption{Distinguished Parameter Regimes from strong stratification (Ia) to weak stratification (IIIc). $*=$boundary regimes.\hfil\break}\label{fig:Regimes}
	\end{center}
\end{figure}
%


\subsection{Geostrophy and the Proudman-Taylor constraint}\label{sec:TP_relax}

The Proudman-Taylor constraint on rapidly rotating fluids arises from the curl of the nondimensional momentum equations (\ref{eqn:BOUS_b_rescaled}a)
\begin{equation}
D_{t}\mbf{\omega} =\left(\mbf{\omega}+\frac{1}{\Ro}\hat{\mbf{z}}\right)\cdot\nabla\mbf{u} +\nabla\times b \hat{\mbf{z}} + \frac{1}{Re_f} \nabla^{2} \mbf{\omega}
\end{equation}
where $\mbf{\omega}=\nabla\times\mbf{u}$ \citep{Proudman_1916,Taylor_1923,Greenspan}.
The leading-order balance at small Rossby numbers is simply $\pd{z}\mbf{u}=0$.
This leading-order balance can be broken if any of the remaining terms in the vorticity equation rise to order $Ro_f^{-1}$; for example, the Proudman-Taylor constraint can be broken in thin viscous boundary layers.
A more relevant example is the quasigeostrophic regime, where the curl of the buoyancy force is sufficiently large to result in the thermal-wind balance at leading order
\begin{equation}
-\partial_z\mbf{u} = \nabla\times b\hat{\mbf{z}}.
\end{equation}

The leading-order asymptotic balance associated with the Proudman-Taylor constraint can be more usefully written as
\begin{equation}
\partial_z\mbf{u} = \mcl{O}(Ro_f)
\end{equation}
which suggests that the Proudman-Taylor constraint can be broken by allowing variation in the $\hat{\mbf{z}}$ direction on vertical scales order $Ro_f^{-1}$ larger than the horizontal scales of motion.
In summary, for weak buoyancy, associated with weak stratification, one expects the dynamics to be tall and thin, whereas for strong buoyancy and strong stratification one expects the dynamics to display an order-one aspect ratio.
The equations used in our the numerical experiments (described in \S \ref{sec:deriv_redeqns}) are geostrophically-balanced, yet break the Proudman-Taylor constraint at small horizontal scales by allowing long vertical variations.
The equations also allow the Proudman-Taylor constraint to be broken on unit-aspect-ratio scales in the presence of sufficiently strong buoyancy forcing.

\subsection{Eddy-wave dispersion relation at $\Ro \ll1$}\label{sec:disp_rel}

A linear analysis of the unforced and inviscid form of equations ($\ref{eqn:BOUS_b_rescaled}$),
for normal modes $\propto \exp[i(\omega t + \mbf{k}_\perp \cdot \mbf{x}_\perp + k_z z) ]$,  provides the inertia-gravity
dispersion relation for the wave frequency of oscillation $\omega$ and the horizontal and vertical
wavenumbers $\mbf{k}_\perp, k_z$: 
\begin{equation}
\omega^{2}_{\mbox{wave}} = \frac{1}{\Fr^{2}}\sin^{2}\theta + \frac{1}{\Ro^{2}}\cos^{2}\theta, \quad
\omega^{2}_{\mbox{eddy}}= 0.
\label{eqn:disp_AZ}
\end{equation}
\noindent Here $\theta = \tan^{-1}(k_{\perp}/k_{z})$ denotes the angle made with the positive $\mbf{z}$-axis.
The dispersion relation (\ref{eqn:disp_AZ}) implies the following bound on the wave frequencies
\begin{equation}
\omega_{\mbox{wave}} \geq \min \left (  \frac{1}{\Fr}, \frac{1}{\Ro}\right).
\end{equation}

It is particularly interesting to interpret the wave dispersion relation in the $\Ro\ll1$
limit as a function of stratification which, as established in the previous section, is tied to the spatial anisotropy of the flow. 
In the presence of strong stratification where $\Fr\ll1$, the wave dispersion relation implies $\omega_{\mbox{wave}}\gg\mcl{O}(1)$ for all $\theta$.
Hence wave and eddy turnover timescales are asymptotically separated for all waves. This is the classical quasi-geostrophic limit where
it is well-established that fast inertia-gravity waves may be filtered from the Boussinesq equations. This reduction  leads to the hydrostatic QG equations describing the evolution of eddies on a slow manifold. 

For weakly stratified flows characterized by $\Fr=\mathcal{O}(1)$ there are fast waves and slow waves, depending on the anisotropy of the wave, i.e.~$\theta$.
The dispersion relation (\ref{eqn:disp_AZ}) clearly shows that waves with $\theta\sim\pm(\pi/2 - \mcl{O}(\Ro))$ retain order-one frequencies in the limit $\Ro\ll1$.
Waves with angle $\theta\sim\pm(\pi/2 - \mcl{O}(\Ro))$ have $k_\perp/k_z\sim\Ro^{-1}$, i.e.~longer vertical than horizontal scales.
It is now seen that these anisotropic inertia-gravity waves are not fast compared to the nonlinear eddy dynamics; since there is no gap between the time scale of waves and the time scale of eddies, the idea of a slow manifold is no longer applicable. 

An approximate dispersion relation for these slow waves is obtained by inserting $k_\perp/k_z\sim\Ro^{-1}$ into the dispersion relation (\ref{eqn:disp_AZ}) and eliminating small terms; the result is
\begin{equation}
\omega^{2}_{\mbox{wave}} \sim \frac{1}{\Fr^2} + \left(\frac{k_z}{k_\perp\Ro}\right)^2,\quad \frac{k_z}{k_\perp}\sim\Ro.\label{eqn:disp_AZ_2}
\end{equation}
The phase and group velocities $\mbf{v}_{\mbox{p}}$
and $\mbf{v}_{\mbox{g}}$ associated with these slow waves are given by 
\begin{subeqnarray}
\mbf{v}_{\mbox{p}} &\sim& \frac{\omega_{\mbox{wave}}}{k_\perp^{2}} \left ( k_{x}, k_{y}, k_{z}\right )=\mcl{O}(1,1,\Ro), \\
\mbf{v}_{\mbox{g}} &\sim&  \left(\frac{k_z}{k_\perp\Ro}\right)^2 \frac{1}{k_\perp^2\omega_{\mbox{wave}}}   \left (- k_{x}, -k_{y},  \frac{k_{\perp}^{2}}{k_{z}} \right )=\mcl{O}(1,1,\Ro^{-1}) 
\end{subeqnarray}
with  $\mbf{v}_{\mbox{p}}\cdot \mbf{v}_{\mbox{g}} = 0$ and $\vert \mbf{v}_{\mbox{g}}\vert  \gg \vert \mbf{v}_{\mbox{p}} \vert$.
Hence, inertia-gravity waves have phase and group velocities that are perpendicular: 
the slow waves propagate predominantly in horizontal
directions whilst wave-energy propagated by the group velocity is transmitted predominantly in the vertical direction \citep{Greenspan}.  We note that
velocity magnitudes are such that information is transmitted on the $\mcl{O}(1)$ eddy-turnover time in all directions; this follows from the fact that  information in the horizontal propagates over $\mcl{O}(1)$ horizontal scales while information in the vertical propagates over $\mcl{O}(\Ro^{-1})$ vertical scales. 
The consequences of wave-eddy interactions without a time scale separation are still not fully understood, primarily
because the main approach has been the use of DNS where efficiency and accuracy becomes increasingly prohibitive
in the $Ro\ll1$ limit. In the following, we analyze, and simulate reduced equations that describe the nonlinear interactions of vortical modes and slow inertia-gravity waves.


\section{Reduced NonHydrostatic QG equations}\label{sec:deriv_redeqns}

Detailed derivations of the NH-QG equations have been documented elsewhere \citep{JKMW_JFM_2006,JFM_SJKW2006,kJ07}. 
In the following, we present the NH-QG equations highlighting only the most salient points.
Deduction of the NH-QG equations proceeds by identifying the Rossby number $\varepsilon \equiv \Ro \ll 1$
as the small parameter and introducing the asymptotic series expansion 
\begin{equation}
\mbf{v} = (\mbf{u},p, b)^T =\varepsilon^{-1} \mbf{v}_{-1} +  \mbf{v}_{0} + \varepsilon \mbf{v}_{1} + \varepsilon^{2} \mbf{v}_{3} + \mcl{O}(\varepsilon^{3})
\label{eqn:asymp_series}
\end{equation}
together with a multiple time scale expansion and a rescaled, anisotropic vertical coordinate 
\begin{equation}
\pd{z} \rightarrow \varepsilon \pd{Z}, \qquad \pd{t} \rightarrow \pd{t} + \varepsilon^2\pd{T}
\label{eqn:scale_expan}
\end{equation}
into the Boussinesq equations. The large vertical scale is precisely the scale at which deviations from the Proudman-Taylor constraint are allowed. The
slow dimensional time scale $T^{*}$ is the period over which the vertical buoyancy flux acts to modify the mean buoyancy profile, and is such that the ratio of the order-one time scale $T^{*}_{f}$ to the slow time scale $T^{*}$ is given by $A_{T} = T^{*}_{f}/T^{*} = \veps^{2}$.
This procedure results in an ordered hierarchy of equation balances that must be solved in succession.
The multiple scales approach requires the following decomposition of each fluid variable into 
mean and fluctuating components, i.e., 
\begin{equation}
\mbf{v} (\mbf{x},Z,t,T) = \ov{\mbf{v}}(Z,T) + \mbf{v}'(\mbf{x},Z,t,T),
\label{eqn:mean_decomp}
\end{equation}
where overbars denote small scale and fast time averages such that 
\begin{equation}
\ov{\mbf{v}}(Z,T) \equiv \frac{1}{\tau V} \int_{\tau,V} f(\mbf{x},Z,t,T) \mathrm{d}\mbf{x}\mathrm{d}t, \quad \ov{\mbf{v}^\prime}\equiv 0.
\label{eqn:mean_defn}
\end{equation}

The non-dimensional parameters and their distinguished relations to $\varepsilon$
are now determined as  \citep{JKMW_JFM_2006} 
\begin{equation}
\Fr = \mcl{O}(1),\quad 
\quad \Eu \sim \varepsilon^{-1},\quad
Re_f = \mcl{O}(1).
\end{equation}
The Reynolds number in particular has an upper bound value $Re_f = o(\Ro^{-2})$ that indicates fluid motions may be driven from 
laminar through to turbulent motions. Importantly, $\Fr$ serves as a control parameter that may be varied from the strong stratification
regime ($\Fr\rightarrow0$) through to the pure rotation regime ($\Fr\gg1$).

An asymptotic perturbation analysis then reveals  $\ov{\mbf{u}}_{-1}=\mbf{v}^\prime_{-1}\equiv \mbf{0}$ together with a leading order mean hydrostatic balance, i.e., 
\begin{equation}
\pd{Z} \ov{p}_{-1}= \ov{b}_{-1}.
\label{eqn:hydrostatic}
\end{equation} 
The leading order dynamics captured by the NH-QG equations are  found to be in pointwise geostrophic balance satisfying 
\begin{subeqnarray}
\hat{\mbf{z}} \times \mbf{u}_0+ \nabla p_0' &=& 0, \\
\nabla \cdot \mbf{u}_0 &=& 0.
\label{eqn:order_0} 
\end{subeqnarray}
This yields, on defining $\nabla_\perp = (\pd{x},\pd{y} ,0)$, the diagnostic solution
 \begin{equation}
 \mbf{u}'_0 = -\nabla_\perp \times  \psi_0' \mbf{\hat{z}}  + w_0^\prime \mbf{\hat{z}}, \qquad p'_0=\psi'_0, \qquad  \ov{\mbf{u}}_0=0.
 \end{equation}

The reduced NH-QG equations describing the flow evolution are deduced at the next order by application of asymptotic solvability conditions and are given by  (dropping primes)
\begin{subeqnarray}
\pd{t}\zeta_0 + J[\psi_0,\zeta_0]-\pd{Z}w_0 &=& \frac{1}{Re_f}\nabla_{\perp}^2\zeta_0,  \\
\pd{t}w_0+J[\psi_0,w_0]+\pd{Z}\psi_0 &=& {b_0}+ \frac{1}{Re_f}\nabla_{\perp}^2w_0 + f_{w_{0}},   \\
\pd{t}b_0+J[\psi_0,b_0]+w_0\left(\pd{Z}\ov{b}_{-1}+\frac{1}{\Fr^2} S(Z) \right)&=&\frac{1}{\Pra Re_f}\nabla_{\perp}^2b_0,\\ 
\pd{T}\ov{b}_{-1}+\pd{Z}\left(\ov{w_0 b_0}\right)&=&\frac{1}{\Pra Re_f}\pd{Z}^2\ov{b}_{-1},
\label{eqn:res_red_eqns} 
\end{subeqnarray}
defining the evolution of vertical vorticity $\zeta_0 = \nabla^2_\perp \psi_0$, vertical velocity $w_0$, and buoyancy
$\ov{b}_{-1}+Ro_f b_0$ decomposed into its mean and fluctuating components.

The NH-QG equations bear the characteristic hallmark of QG theory, namely:  $p'_0=\psi_0$ serves as the geostrophic stream function;
planetary rotation is solely responsible for axial vortex stretching in  equation (\ref{eqn:res_red_eqns}a); material advection occurs solely in the horizontal direction with $ \mbf{u}_{0\perp}\cdot\nabla_{\perp}\equiv J[\psi_0,\cdot] =\pd{x}\psi_0\pd{y}-\pd{y}\psi_0\pd{x} $, 
vertical advection is a subdominant phenomenon with $w'_0 \pd{Z}\mbf{v}'_0= \mcl{O}(\varepsilon)$. 
However, in the presence of weak stratification, vertical motions are now significant and result in the appearance of inertial acceleration terms in vertical momentum equation  (\ref{eqn:res_red_eqns}b). Notably, linearization about a constant stratification profile $S(Z)=1$ in the inviscid limit
$Re_f\rightarrow \infty$ captures the dispersion relation  for both slow inertial-gravity waves (\ref{eqn:disp_AZ_2}) and eddies $\omega_{\mathrm{eddy}}=0$. The NH-QG equations thus reflect the fact that slow inertial-gravity waves and eddies interact nonlinearly in the rapidly rotating, weakly stratified regime.

\subsection{Energetics and conserved quantities}\label{sec:Conserved}
Like the Boussinesq equations, the inviscid and unforced NH-QG equations conserve several positive quadratic functionals.
The time-rate-of-change of horizontal kinetic (HKE), vertical kinetic energy (VKE) and potential energy (PE) are given\footnote{`Potential energy' here is not an approximation to the gravitational potential energy $-g\la\ov{\rho z}^\mcl{A}\ra$, but the terminology is conventional.}, respectively, by 
\begin{subeqnarray}
\pd{t}\mbox{HKE} &:=& \pd{t}\left [\frac{1}{2}\left(\la \ov{|\nabla_{\perp}\psi_0|^{2}}^\mcl{A} \ra  \right)\right ]
= \la \ov{w_0 \pd{Z}\psi_0}^\mcl{A}  \ra, \\
\pd{t}\mbox{VKE} &:=& \pd{t}\left [\frac{1}{2}\la \ov{w_0^{2}}^\mcl{A} \ra \right ]
= -\la \ov{w_0 \pd{Z}\psi_0}^\mcl{A}  \ra + \la \ov{w_0b_0}^\mcl{A}  \ra, \\
\pd{t}\mbox{PE} &:=&  \pd{t} \left [\frac{1}{2} \left\la \frac{\ov{b_0^{2}}^\mcl{A} }{\left (\pd{Z}\ov{b}_{-1}(Z)+\Fr^{-2}S(Z)\right )} \right \ra \right ]
=-{\la \ov{w_0b_0}^\mcl{A}  \ra},
\label{eqn:KEPE}
\end{subeqnarray} 
where $\la \cdot \ra$ and $\ov{\cdot}^\mcl{A}$ denote vertical and horizontal averages, respectively, and 
the time-invariance of total energy $E = KE + PE = \mbox{HKE} + \mbox{VKE} + \mbox{PE}$ is clear. 
The equations also conserve a total buoyancy variance 
\begin{equation}
\pd{t}\la \ov{(b_0^2+(\ov{b}_{-1}+\Sigma(Z))^2}^\mcl{A}\ra=0,\quad S(Z) := \pd{Z}\Sigma(Z) = -\Fr^{-2}\pd{Z}\delta\hat{\rho}.
\end{equation}
Finally, the NH-QG equations materially conserve a form of potential vorticity (PV)
\begin{subeqnarray}
&\pd{t}q + J[\psi_0,q ] = 0, \\ 
&q = \zeta_0 + \left(\mbf{\omega}_{\perp}\cdot\nabla_{\perp} + \pd{Z} \right) \left(\frac{b_0}{\left (\pd{Z}\ov{b}_{-1}+\Fr^{-2}S(Z)\right )}  \right ). 
\end{subeqnarray}
Notably, it can be seen the potential vorticity $q$ can be partitioned into a linear and nonlinear component dependent on vortical and vertical motions respectively. 

\subsection{Barotropic, baroclinic decompostion}\label{sec:BTBC}

Rapid rotation often induces a transfer of energy to the depth-independent component of horizontal velocity \citep{SW_PoF_1999}.
It is useful therefore to examine the energetic interaction of the depth-independent horizontal velocity with the remainder of the system.
In quasigeostrophic theory, the velocity is often expanded as a sum over a basis of vertical modes, the first of which is depth-independent and is conventionally called the `barotropic' mode \citep{RYG16}.
More generally, the definition of a barotropic fluid is a fluid for which density is a function of pressure alone.
A constant-density fluid is an example of a barotropic fluid, but a constant-density fluid need not be depth-independent -- an apparent conflict with the conventional quasigeostrophic usage of the term.

To fix a particular usage of the terms `baroclinic' and `barotropic' in the context of a stratified Boussinesq fluid we take the following line of reasoning.
In a Boussinesq fluid the deviation from the constant reference density is $-b^{*}\rho_0^{*}/g$, which is not generally a function of pressure alone unless $b^{*}=0$.
Because vertical velocity in the presence of a backckground stratification induces buoyancy perturbations, $w$ is intimately associated with baroclinicity and we choose to consider it as part of the `baroclinic' component of the dynamics.
The barotropic component, having both $b_0=0$ and $w_0=0$, must also have no vertical pressure gradient $\pd{Z}\psi_0=0$.
This line of reasoning leaves the depth-independent part of the horizontal velocity as the only element of the barotropic component, with the baroclinic component comprising $w_0$, $b_0$, and the depth-dependent part of $\psi_0$.
Our use of the terms is distinguished from an alternate use where `barotropic' simply indicates the depth-independent component and includes both $\la w_0\ra$ and $\la b_0\ra$.

We thus arrive at the barotropic-baroclinic (bt-bc) decomposition
\begin{eqnarray}
&\mbf{u}_{0,bt} = -\nabla_\perp \times \la \psi_0 \ra \mbf{\hat{z}},\hspace{2em} \quad & b_{0,bt}= 0,\\
& \mbf{u}_{0,bc} = -\nabla_\perp \times  \psi'_0  \mbf{\hat{z}} + w_0 \mbf{\hat{z}} , \quad & b_{0,bc} = b_0 \nonumber
\end{eqnarray}
where $\psi_0 =  \la \psi_0 \ra + \psi'_0$. Partitioning the NH-QG equations thus reduces to decomposing the vorticity equation (\ref{eqn:res_red_eqns}a), into its barotopic and baroclinic components. Namely
\begin{subeqnarray}
\pd{t}\la \zeta_0 \ra + J[\la \psi_0 \ra,\la \zeta_0 \ra] = -\la J[\psi'_0,\zeta'_0] \ra +  \frac{1}{Re_f}\nabla_{\perp}^2\la \zeta_0 \ra, \\
\pd{t}\zeta'_0 + J[\la \psi_0 \ra + \psi'_0, \zeta'_0] -\la J[ \psi'_0, \zeta'_0] \ra  + J[\psi'_0,\la \zeta_0 \ra]  - \pd{Z}w'_0 = \frac{1}{Re_f}\nabla_{\perp}^{2}\zeta'_0.
\label{eqn:BT_BC}
\end{subeqnarray} 
\noindent Equation (\ref{eqn:BT_BC}$a)$
is the two-dimensional barotropic vorticity equation. Within the barotropic subspace kinetic energy \
$\ov{\vert \nabla_\perp \la \psi_0 \ra \vert^2}^\mcl{A}$ and 
enstrophy $\ov{\la\zeta_0\ra^2}^\mcl{A}$ are conserved quantities in the absence of dissipation and forcing. Forcing of barotropic vorticity 
occurs through nonlinear interactions between purely baroclinic fields in the form of advection of baroclinic vorticity by baroclinic
horizontal velocities, i.e., $\la J[\psi'_0,\zeta'_0] \ra = \la \mbf{u}_{0\perp}' \cdot \nabla \zeta'_0\ra$. Therefore, this term
acts as a source when $\mbf{u}_{0\perp}'$ and $\nabla \zeta'_0$ are barotropically collinear. 

Some comments are appropriate on the distinguishing features of the NH-QG equations in comparison with a recent and alternative formulation by 
\citet{WEHT_JFM_2011}. In \citet{WEHT_JFM_2011} the asymptotic development is based strictly on a multiple-scales approach in time only with an
isotropic scaling of the spatial coordinates.  The resulting slow manifold is found to be one that strictly enforces the Proudman-Taylor constraint of the
velocity field, i.e., $\pd{Z} \mbf{u}_0= 0$. 
Consequently, the term coupling baroclinic and barotropic dynamics $\la J[\psi'_0,\zeta'_0] \ra$ is predicted to be asymptotically small, therefore decoupling
barotropic vorticity dynamics from the now Taylorized depth-independent baroclinic dynamics of $\la w_0 \ra$ and $\la b_0 \ra$. 
Stochastically forcing baroclinic dynamics therefore cannot influence barotropic motions \citep{WW_JFM_2014}. We contend that the NH-QG equations
 demonstrate that slow inertial-gravity waves and baroclinic eddies are a vital leading-order component of the dynamics at low Rossby and moderate
 Froude numbers. 


\section{Numerical simulation for stably stratified NH-QG equations}\label{sec:num_meth}

\begin{table}
\begin{center}
\def~{\hphantom{0}}
\setlength\tabcolsep{.3cm}
\begin{tabular}{cc|ccc}
Regime &$\Fr(Re_{f})$ & $\Fr(Re_{f}=50)$ & $\Fr(Re_{f}=100)$ & $\Fr(Re_{f}=300)$ \\
\hline
Ia    & $\frac{1}{2}Re^{-1/2}$ & 0.0707 & 0.0500 & 0.0289 \\
Ib*  & $Re^{-1/2}$ & 0.1414 & 0.1000 &  0.0577 \\
Ic    &	$\frac{1}{2}(1+Re^{-1/2})$ & 0.5707 & 0.5500 & 0.5289 \\
II*   &	$1$	& 1  & 1 &  1 \\
IIIa  & $\frac{1}{2}(1+Re^{3/4})$ & 9.9015 & 16.311 &  36.542 \\
IIIb* & $Re^{3/4}$ & 18.803 & 31.623	& 72.084 \\
IIIc   & $2 Re^{3/4}$ & 37.606 & 63.246 & 144.17 \\
\hline
Grid resolution &  $N_{x}\times N_{y}\times N_{z}$  & $96\times96\times96$ & $192\times192\times192$  & $384\times384\times384$ \\
\end{tabular}
\end{center}
\caption{Values of $\Fr$ as a function of $Re_{f}$ used in simulations of the NH-QG equations based on the seven regimes identified in figure \ref{fig:Regimes}. Domain size for each simulation is $10L_{f} \times 10L_{f}\times 1$, where $L_{f}=1 $ is the imposed forcing length scale. To ensure sufficient resolution we use the convention that $\Delta x = 2L_{K}$, where $L_{K} = Re^{-3/4}$ is the dissipation length scale, giving the number of Fourier modes used in each Cartesian direction as $N_{x,y,z} = L_{b}Re^{3/4}/2$. The Prandtl number is fixed at $\Pra = 7$ for all simulations.}
\label{tab:parameters}
\end{table}%

Since the layer of stably stratified fluid is void of a natural instability capable of inducing fluid motion, artificial forcing
is required. Previous studies have accomplished the task of forcing a stable layer through the
controlled injection of motion inducing energy \citep{SW_JFM_2002,L_JFM_2006,WEHT_JFM_2011}. The
present study induces fluid motions in a fashion similar to these past investigations. In particular, we perfom
numerical simulations where motion is induced by a controlled injection of vertical kinetic energy. In 
forcing the vertical momentum equation only this study differs from those in which all three components of momentum are
forced \citep[e.g.~][]{SMRP_PRE_2012,MMRP_EPL_2013}, however, by only forcing vertical velocity the energy is injected only into wave modes.
Therefore, energy transfer to the vortical modes must occur through interactions among these linear eigenmodes.

The energy source occurs through the 
vertical momentum equation (\ref{eqn:res_red_eqns}$b$) where forcing
takes the form of the spatially-correlated, white in time stochastic forcing $f_{w_{0}}$. 
The stochastic 
forcing function $f_{w_{0}}$ has a spherically symmetric spectrum     
\begin{equation}
E_{f_{w_{0}}}(\mbf{k}) = C\eps_{f}\exp \left(-\frac{1}{2}(|\mbf{k}|-k_{f})^{2}\right), 
\end{equation} 
where $\eps_{f}$ is the flux of vertical kinetic energy into the system at forcing wavenumber $k_{f}$. For this
study we set $k_{f} = 2 \pi$ (setting the nondimensional horizontal length scale to $L_{f} = 1$) and
$\eps_{f} = 1$ and we normalize the spectrum of the forcing function so that volume averaged
energy flux becomes
\begin{equation}
\int_{0}^{2 \pi}\int_{0}^{\pi}\int_{0}^{\infty} E_{f_{w_{0}}}^{2}(k) k^{2} \sin{\phi} dk d\phi d\theta = 1.
\end{equation}

Numerical simulations of the NH-QG equations are performed in a periodic box and solutions are expanded in
Fourier series. The numerical box has dimensional size $L_{b} L_{f} \Ro H^{*} \times L_{b} L_{f} \Ro H^{*}
\times H^{*}$, where $L_{f} = 1$ is the nondimensional forcing length scale and $L_{b}$ is the nondimensional length of the
horizontal domain, thus, the nondimensional domain size is $L_{b} \times L_{b} \times 1$. The numerical
time-stepping scheme used is an implicit/explicit formally second-order
Runge-Kutta scheme derived by \citet{SMR_JCP_1991} and previously used by \cite{SJKW_JFM_2006}
for numerical simulation of the NH-QG equations for the rapidly rotating Rayleigh-B\'{e}nard problem. 
The delta-correlated forcing is discontinuous everywhere (in time; it is spatially-smooth) so it cannot be treated with standard numerical methods for deterministic differential equations that assume some level of smoothness.
The stochastic dynamics are here treated with a simple splitting method where the deterministic dynamics are treated independently of the stochastic forcing.
To wit, after completing a full time step of the deterministic dynamics a random forcing increment $\sqrt{dt}\chi(\mbf{x},t)$ is added to the solution for
 $w_{0}$ (or, in some initial tests, to $b_{0}$), effectively using the Euler-Maruyama method on the stochastic forcing term $f_{w_0}(t)$ \citep{Higham01}. 
In addition to respecting the stochastic nature of the dynamics, this approach has the desirable property that the mean rate of energy injection is independent of the system state, and can be controlled a priori.

Fourier expansions are
dealiased using the standard 2/3s rule. To ensure sufficient resolution we use the convention that $\Delta x = 2L_{K}$, where 
$L_{K} = Re^{-3/4}$ is the dissipation length scale for statistically steady flow. Use of this convention gives the
number of Fourier modes used in each Cartesian direction as $N_{x,y,z} = L_{b}Re^{3/4}/2$. Resolutions used
in our numerical simulations are given in table \ref{tab:parameters}. 

The simulation parameters $(Re_{f},\Fr,\Pra)$ are selected based on the regimes identified in figure \ref{fig:Regimes}. 
For a given $Re_{f}$ we vary $\Fr$ so as to explore each of the seven regimes identified in figure \ref{fig:Regimes}.
This process of selecting $\Fr$ is outlined in table \ref{tab:parameters}. All simulations are computed with $\Pra = 7$. 

In addition to forcing vertical velocity we have also performed numerical simulations with buoyancy forcing as in \citet{WW_JFM_2014}, however,
since the momentum equations decouple from the buoyancy equations for large $\Fr$ the injection of potential energy 
becomes unphysical since. 
For this reason we only present results associated with the injection of vertical kinetic energy via the vertical velocity equation (\ref{eqn:res_red_eqns}$b$). 

\section{Results}\label{sec:results}

The nondimensional parameters defined in section~\ref{sec:govn_eqns} are based on a priori characteristic scales built from the energy injection rate $\eps_{f}$
and injection scale $L_{f}$. These scales are not necessarily the same as the scales that truly characterize the flow; certainly it is not the case that 
the large-scale flows observed here occur on the forcing scale $L_{f} = 1$. For this reason we give a summary of a posteriori nondimensional parameters
that define the flows simulated. To do this we compute the centroid of energy spectra to get a characteristic wavenumber $k_{c}$ and associated length scale $L_{c}$; we compute a characteristic velocity $U_{c}$ from the volume-averaged horizontal kinetic energy (HKE), that is,

\begin{equation}
k_{c} = \frac{\int k E(k)dk}{\int E(k)dk}, \quad U_{c}=(2HKE)^{1/2},
\label{eqn:centroid}
\end{equation}  

\noindent where $E(k)$, for example, are the curves in figure~\ref{fig:spectra}. These nondimensional measured values are then used to define a posteriori Reynolds and
Froude numbers 

\begin{equation}
Re_{c} = \frac{U_c^*L_c^*}{\nu} = \frac{U_{f}^*U_{c}L_{f}^*L_{c}}{\nu} = Re_{f}U_{c}L_{c}, \quad
Fr_{c} = \frac{U_c^*}{N_0^*L_c^*} = \frac{U_{f}^*U_{c}}{N_0^* L_{f}^*L_{c}} = \Fr \frac{U_{c}}{ L_{c}}  
\end{equation}

\begin{figure}
	\begin{center}
		\includegraphics[width=0.75\linewidth]{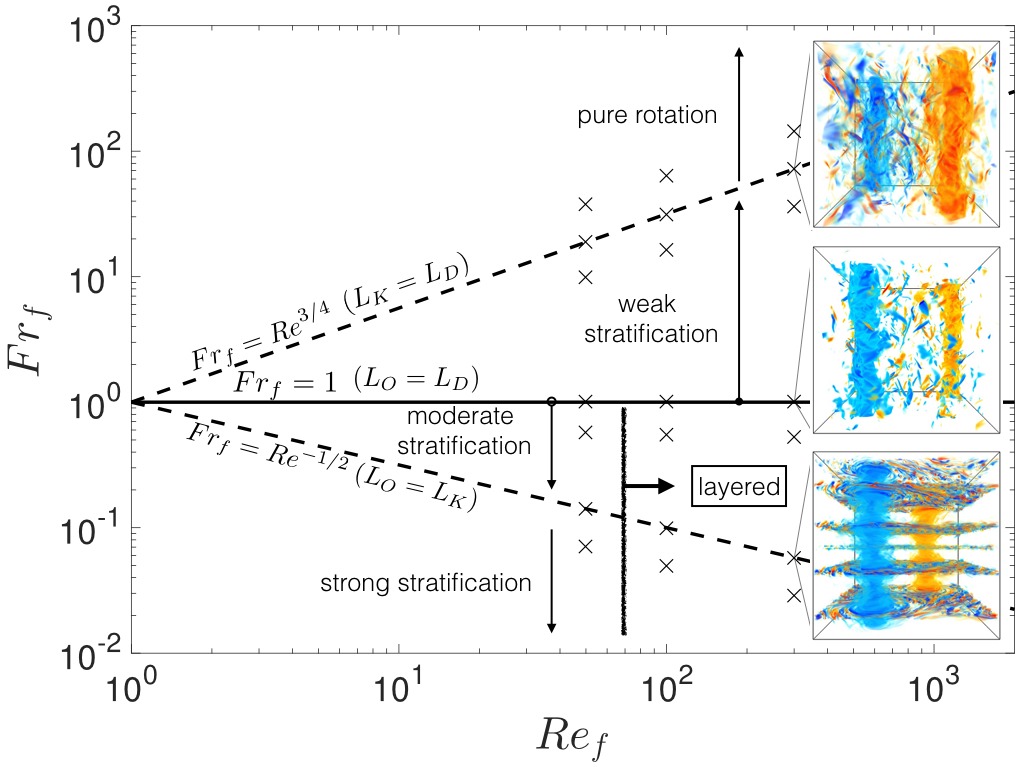} 
		\caption{A qualitative partitioning of $(Re_f,\Fr)$-space using volume renders of vorticity.  Values of $(Re_{f},\Fr)$ for which 
		simulations were performed are denoted by an $\times$ (see Table~\ref{tab:parameters}). 
		The flow is characterized by layering, barotropization and an inverse cascade.  For $\Fr < 1$ the flow organizes into
		well-defined layers (except at low-$Re_{f}$, e.g., $Re_{f} =50$) and when $\Fr \geq 1$ layering is absent.
		We emphasize the presence of a dominant barotropic component of energy and a clear inverse cascade for all $\Fr$ simulated.
		Similar flow characteristics are observed for buoyancy and vertical velocity (see Figures~\ref{fig:flow_wb_strong} and 
		\ref{fig:flow_wb_weak}).}\label{fig:regime_diagram}
	\end{center}
\end{figure}

\noindent A posteriori $Fr_{c}$ and $Re_{c}$ for a range of are parameters are
summarized in table~\ref{tab:non_dim_nums}. Generally, characteristic horizontal scales are larger than $L_{f}$, and characteristic velocities are larger than $U_f$.
This results in Reynolds numbers that are larger than $Re_{f}$. The larger measured horizontal scale $L_{c}$ outweighs the increase in $U_c$, leading to Froude numbers that are smaller, in some cases by an order(s)-of-magnitude, than $\Fr$, however, what was considered weakly stratified as measured by $\Fr$ remains so as measured by $Fr_{c}$. \\

\begin{figure}
	\begin{center}
		\includegraphics[width=0.32\linewidth]{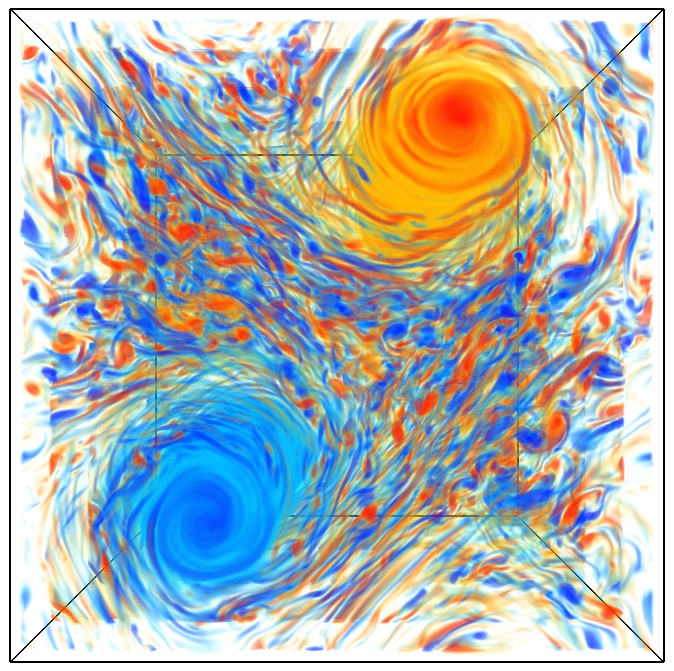} 
		\includegraphics[width=0.32\linewidth]{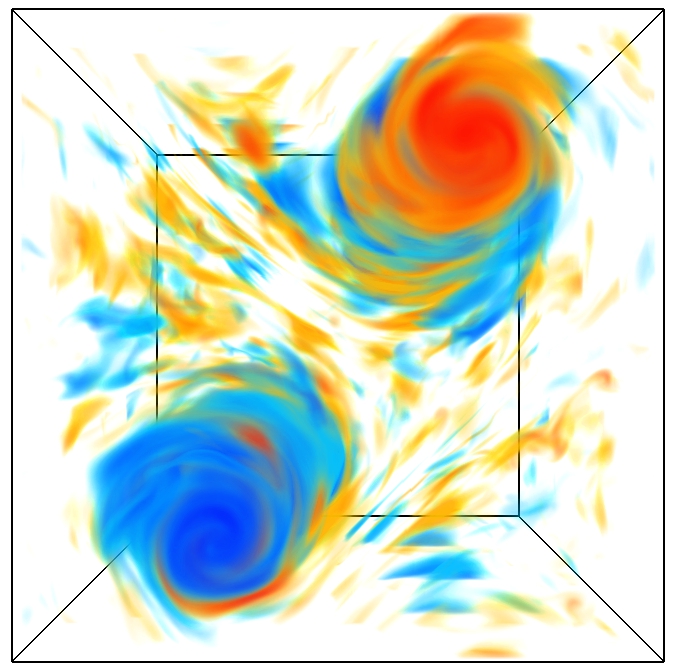} 
		\includegraphics[width=0.32\linewidth]{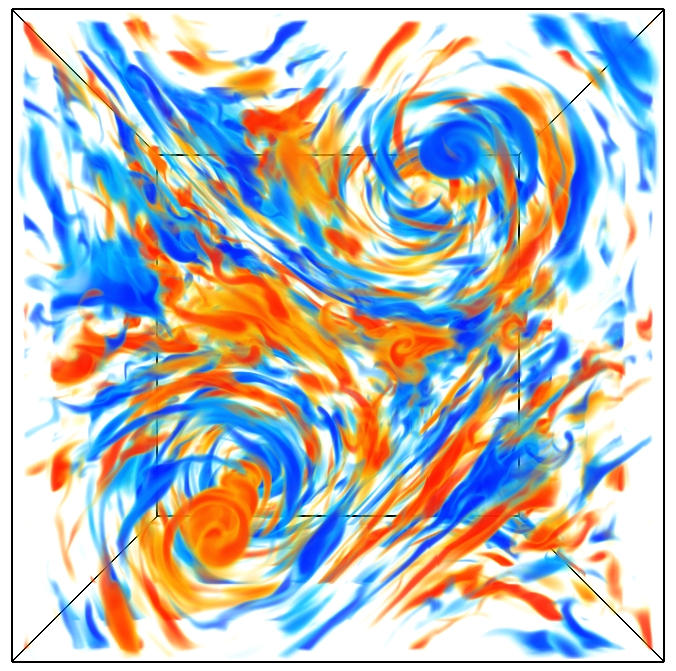} \\ 
		\subfloat[Vertical vorticity, $\zeta$]{\includegraphics[width=0.32\linewidth]{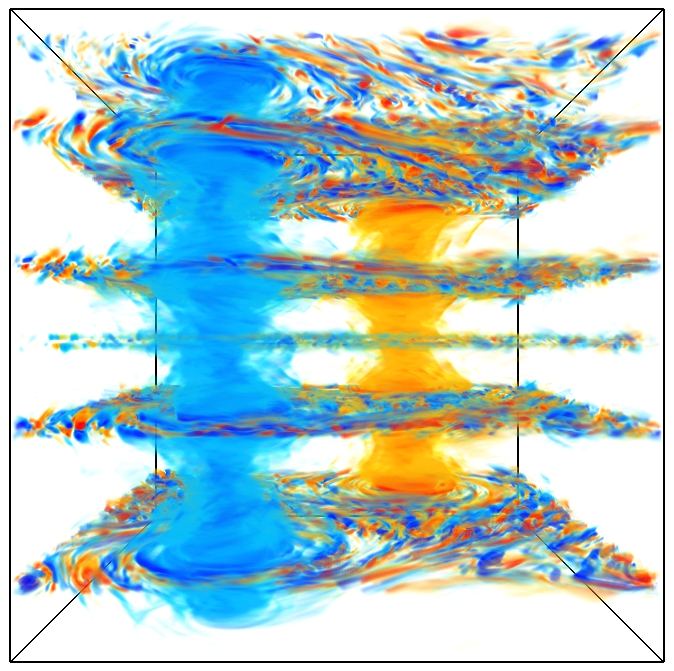}} 
		\subfloat[Buoyancy, $b$]{\includegraphics[width=0.32\linewidth]{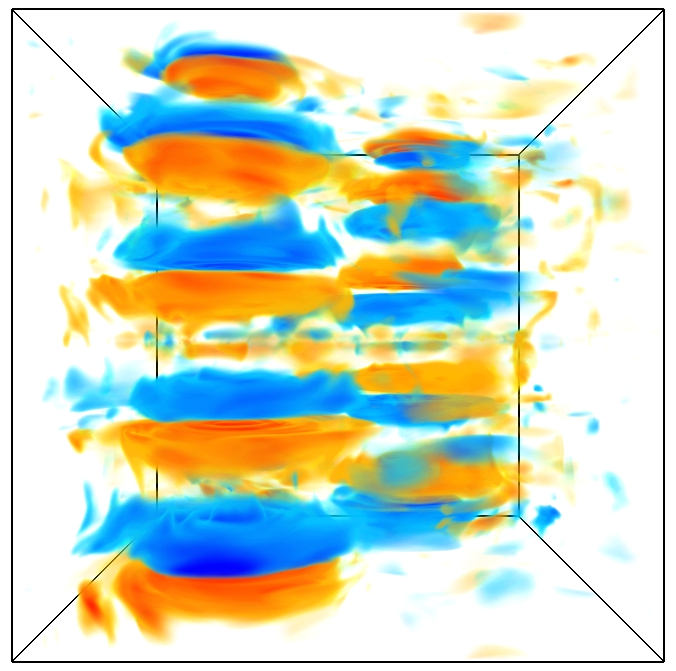}} 
		\subfloat[Vertical velocity, $w$]{\includegraphics[width=0.32\linewidth]{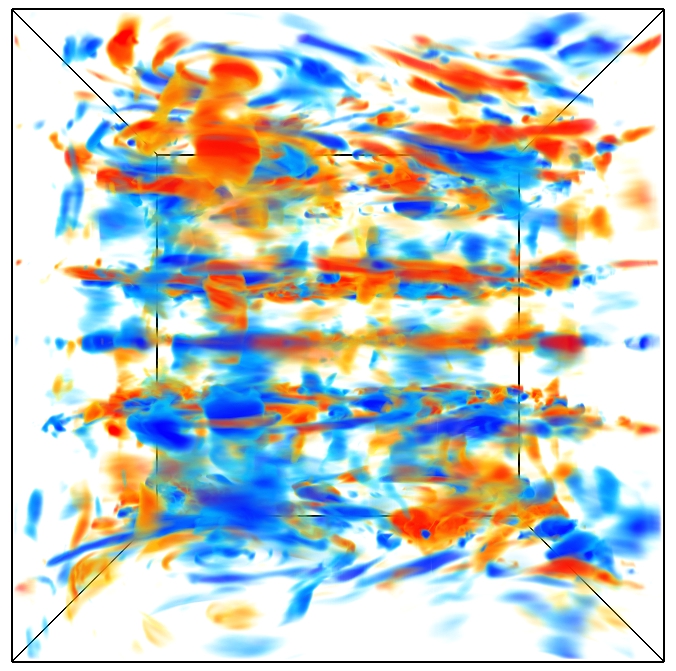}} \\
		\caption{Volume renders of vertical vorticity $\zeta$ (left column), buoyancy $b$  (middle column), and vertical velocity $w$  (right 
		column) for the case of strong stratification $Re_{f} =300,\ \Fr=Re_{f}^{-1/2}$. Top row (top view), bottom row (side view).}
		\label{fig:flow_wb_strong}
	\end{center}
\end{figure}

\begin{figure}
	\begin{center}
		\begin{tabular}{ccc}
		\includegraphics[width=0.32\linewidth]{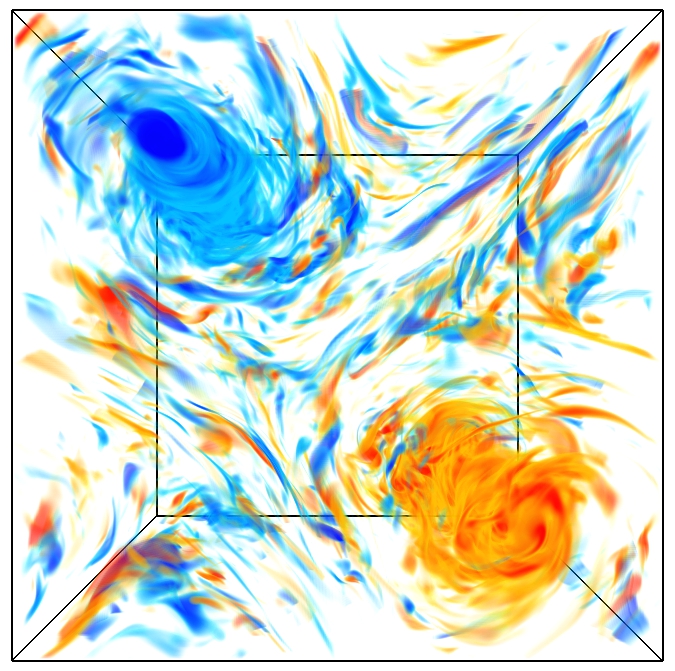} &
		\includegraphics[width=0.32\linewidth]{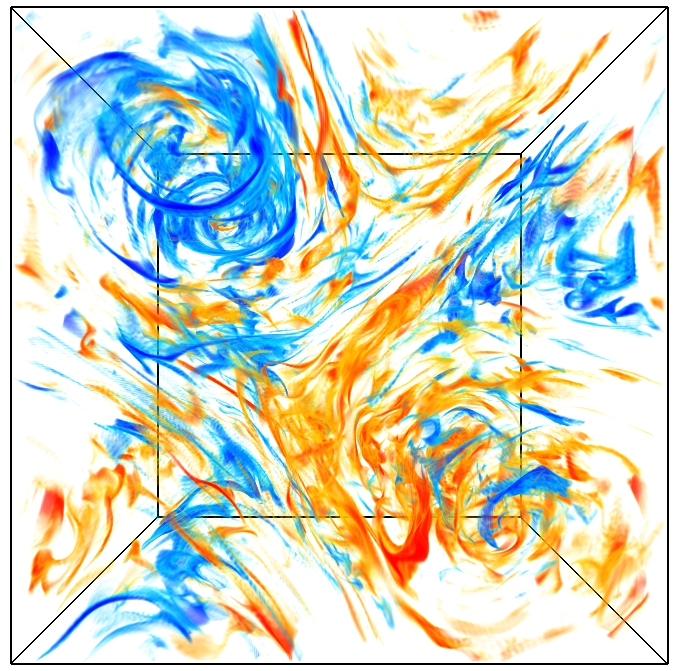} &
		\includegraphics[width=0.32\linewidth]{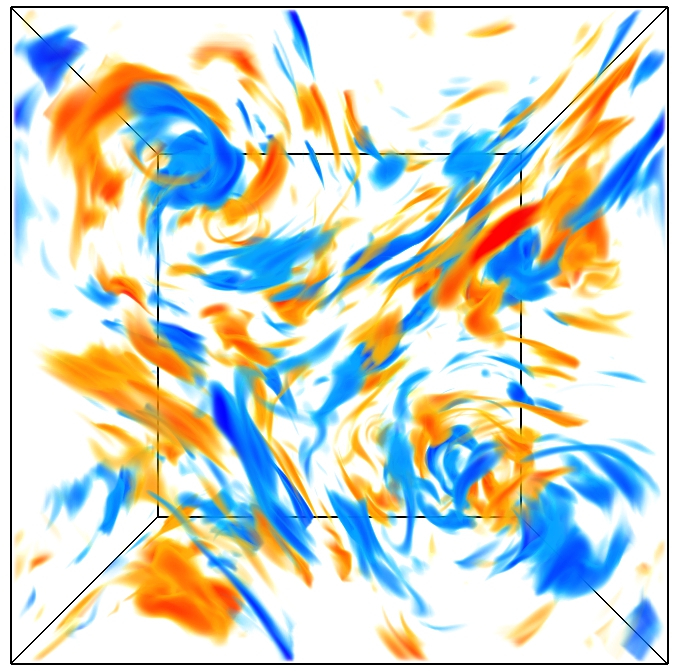} \\ 
		\subfloat[Vertical vorticity, $\zeta$]{\includegraphics[width=0.32\linewidth]{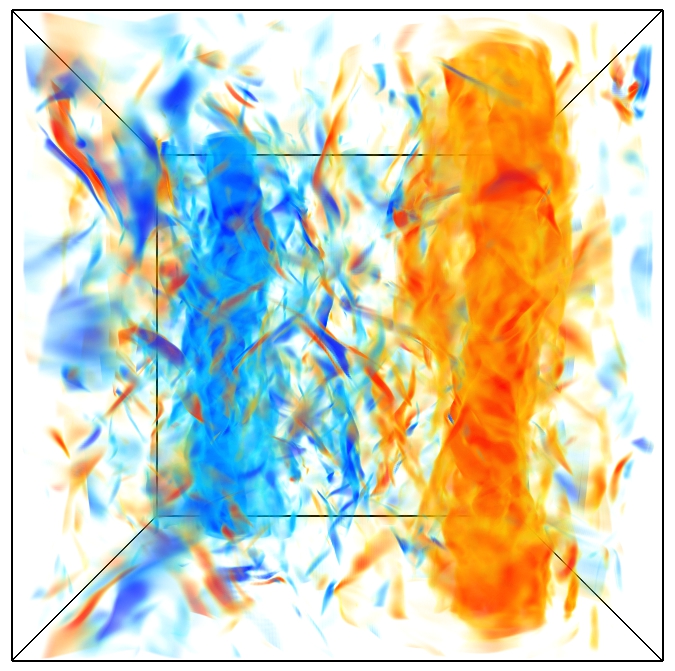}} &
		\subfloat[Buoyancy, $b$]{\includegraphics[width=0.32\linewidth]{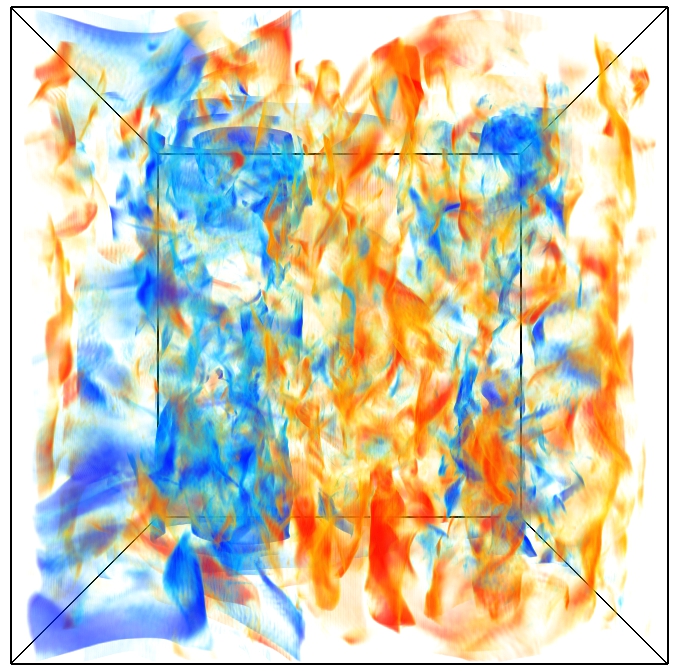}} &
		\subfloat[Vertical velocity, $w$]{\includegraphics[width=0.32\linewidth]{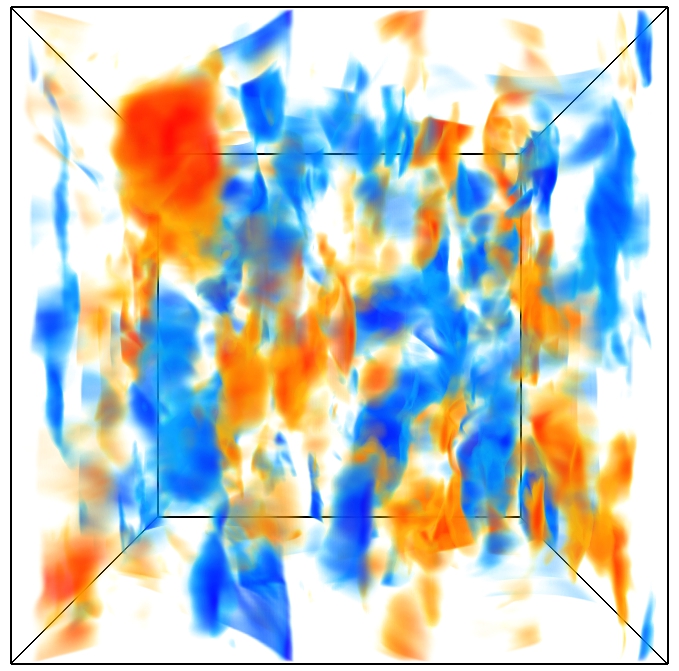}} \\
		\end{tabular}
		\caption{Volume renders of vertical vorticity $\zeta$ (left column), buoyancy $b$  (middle column), and vertical velocity $w$  (right 
		column) for the case of weak stratification $Re_{f} =300,\ \Fr=Re_{f}^{3/4}$. Top row (top view), bottom row (side view).}
		\label{fig:flow_wb_weak}
	\end{center}
\end{figure}

Performing DNS of the NH-QG equations (with the nondimensional parameters outlined in table~\ref{tab:parameters})
two qualitatively identifiable regimes are observed, corresponding to strong and weak stratification: $\Fr<1$ and $\Fr\ge1$ respectively.
The regime diagram in figure~\ref{fig:regime_diagram}
partitions $(Re_f,\Fr)$-space into two regimes based on volume renders of vertical vorticity.
In both regimes the flow organizes into a large-scale, barotropic dipole with some additional small-scale turbulence. Figures~\ref{fig:flow_wb_strong} and \ref{fig:flow_wb_weak} 
gives renders for vertical vorticity, buoyancy and vertical velocity for strong and weak stratifications when $Re_{f} = 300$.

The strong stratification regime ($\Fr < 1$, figure~\ref{fig:flow_wb_strong}) is distinguished by a tendency of the flow to form well-defined and sustained layers where small-scale turbulence is active and the local stratification is reduced.
Layering is observed for $Re_{f} = 100$ and $Re_{f} = 300$, but not for $Re_{f} = 50$.
We conclude that the instability responsible for layering is inhibited by
viscous effects at lower $Re_{f}$. We note that layering, as observed in figure~\ref{fig:flow_wb_strong} is not observed for
classical QG dynamics where energy rapidly transfers to large vertical scales \citep{SV01,SV02}.
In the second regime of weak stratification ($\Fr \geq 1$, figure~\ref{fig:flow_wb_weak}) the columnar structures
are unobstructed by layers, and evolve in a sea of small-scale turbulence. 

\begin{table}
	\begin{center}
	\def~{\hphantom{0}}
		\setlength\tabcolsep{.5cm}
		\begin{tabular}{c c|c c|c c}   
                 $Re_f $&$Re_c $& $\Fr $& $Fr_c $& $L_c $&$ U_c $ \\
                 \hline
		50  & 150  & 0.1414 & 0.0118 & $6.0 $ & 0.5  \\
		100& 980  & 0.1000 & 0.0200 & $7.0 $ & 1.4  \\
		300&4290 & 0.0577 & 0.0195 & $6.5 $ & 2.2  \\
		\hline
		\hline
		50  & 604  & 18.80 & 4.50 & $7.1 $ & 1.7  \\
		100&2190 & 31.62 & 12.99 & $7.3 $ & 3.0  \\
		300&5742 & 72.08 & 31.67 & $6.6 $ & 2.9  \\
		\hline
    		\end{tabular}
	\caption{Characteristic scales $U_{c}$ and $L_{c}$ computed from centroids of energy spectra and nondimensional quantities $Re_{c}$ 
	and $Fr_{c}$ based on the measured values  $L_{c}$ and $U_{c}$. }
	\label{tab:non_dim_nums}
	\end{center}
 \end{table} 

In both regimes the energy accumulates primarily in the barotropic mode and at large horizontal scales, indicating a robust inverse cascade of energy.
At lower Reynolds numbers $Re_{f}\le 100$ ($Re_c$ up to $\approx2000$) the total energy in the system reaches a statistical equilibrium.
In addition to the inverse energy transfer, we diagnose a robust \emph{direct} transfer of kinetic energy in the barotropic mode, which allows the small amount of energy injected by the baroclinic motions to be balanced by small-scale dissipation, leading to energy saturation.
At higher Reynolds numbers, $Re_{f}=300$ ($Re_c$ greater than $\approx4000$), the total energy shows no sign of saturation.
These results are presented in more detail in the following subsections.

We note that these results do not necessarily represent universal properties of rotationally constrained stratified flow in every respect. Undoubtedly,
the dynamic behavior depends significantly on the method by which external energy is injected to excite motion. As
mentioned above, the forcing method employed here excites vertical motion, therefore, only excites wave modes and does not
directly force the vortical mode. This approach to forcing aims to better understand the
energetic pathway from three-dimensional baroclinic motions to two-dimensional barotropic motions.
\subsection{Layering}\label{sec:layering}

\begin{figure}
	\begin{center}
		\begin{tabular}{ccc}
		$(a)$ & $(b)$ & $(c)$ \\
		\includegraphics[width=0.315\linewidth]{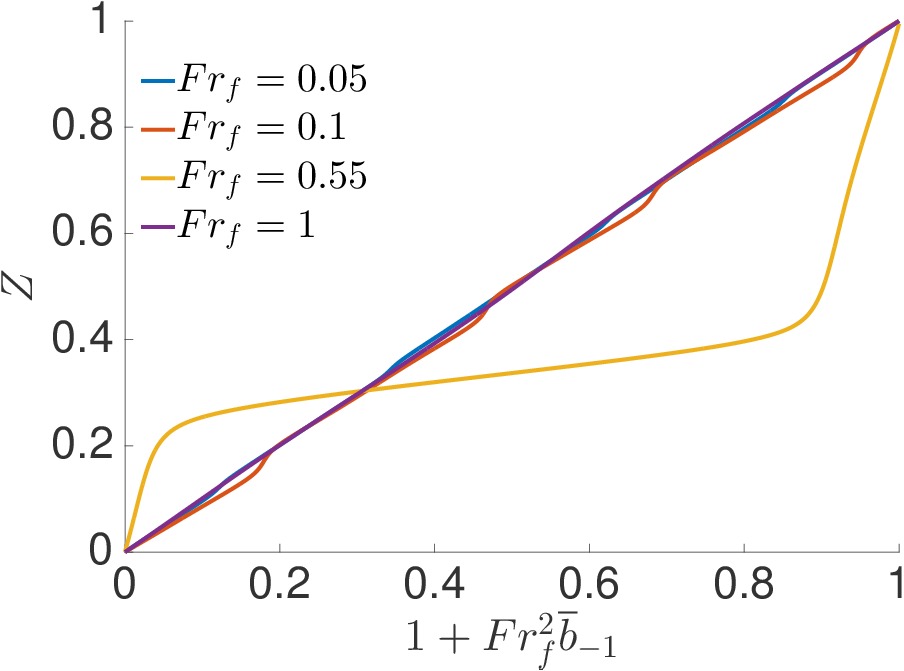} &
		\includegraphics[width=0.315\linewidth]{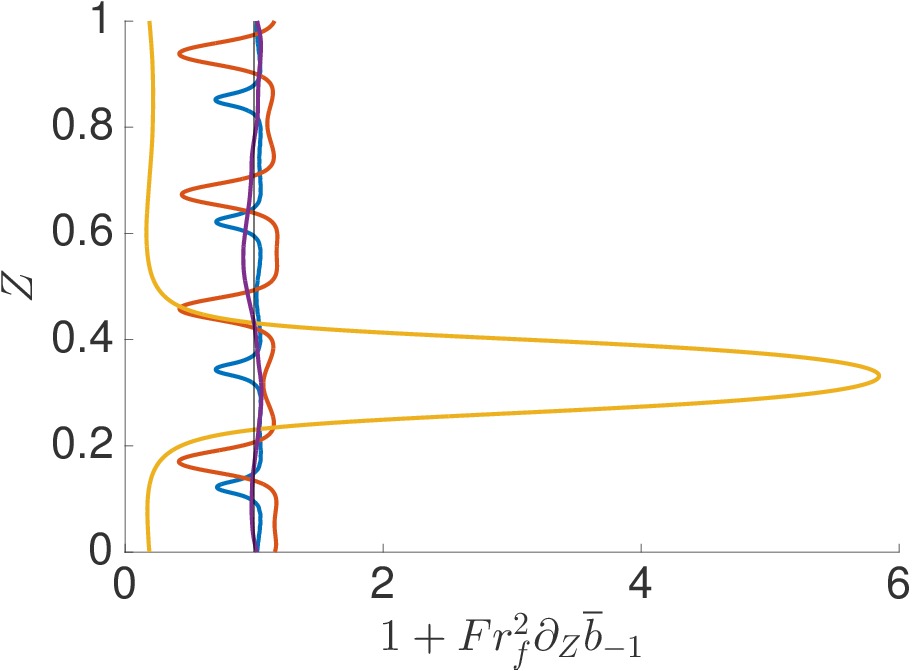} &
		\includegraphics[width=0.315\linewidth]{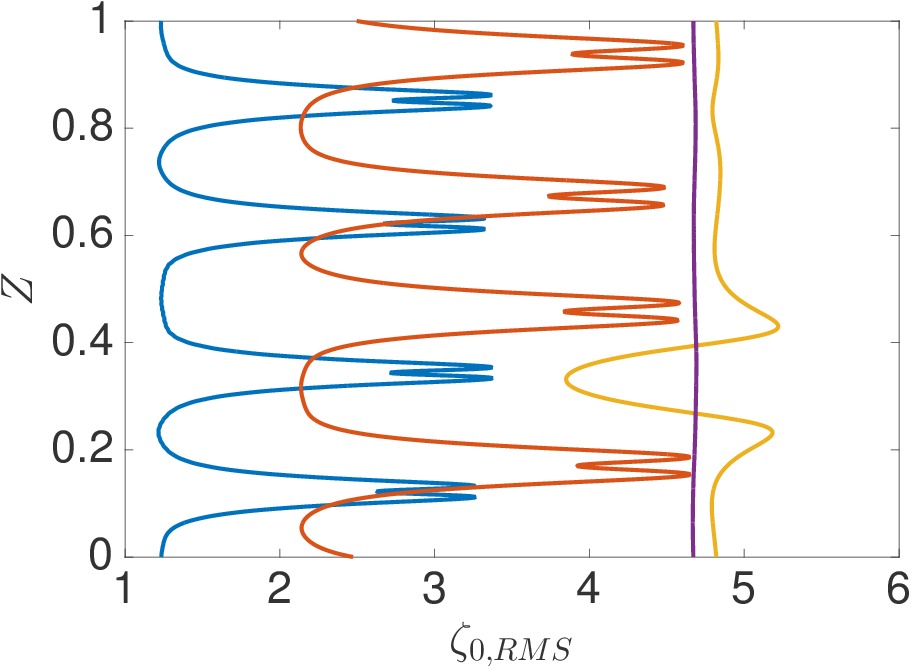} \\
		\end{tabular}
		\caption{Time averaged vertical profiles for $Re_{f} = 100$. Profiles of $(a)$ total mean buoyancy, $(b)$ vertical gradient of mean buoyancy
		and $(c)$ RMS vertical vorticity. Layering occurs in horizontal planes where mean stratification is locally minimized.
		The effect on the stratification profile is due to the nature of vertical buoyancy flux, similarly, layered structuring seen for vertical vorticity
		is due vortex stretching. Layer locations coincide with locations of sharp local minima within the peaks of $\zeta_{0,RMS}$. Layer height
		may be given by the distance between local maxima surrounding the singular local minima and indicate the presence of sublayers (jets). 
		The vertical extent of layers and their sublayers is observed to increase with decreased stratification. 
		Similar structuring is observed for vertical velocity, buoyancy, and dissipation.}
		\label{fig:profile_3}
	\end{center}
\end{figure}

Layering is observed in all fields though most distinct in the renders of vertical vorticity shown in figure~\ref{fig:flow_wb_strong}. To clarify terminology, we define layers to be the localized planar regions home to small-scale turbulence and occurring for $\Fr<1$. 
Figure~\ref{fig:profile_3} shows the effect of strong stratification on the time-averaged vertical gradient of the total mean buoyancy profile and on the structure of  $\zeta_{0,RMS}$ for simulations with $Re_{f} = 100$ and $\Fr \leq 1$. 
Reduction of stratification within the layers is presumably associated with local turbulent mixing within the layers.


Some basic characteristics of the location and height of layers are given by the mean buoyancy gradient and vertical profiles of $\zeta_{0,RMS}$. The more
informative of the two is the set of RMS profiles of vertical vorticity. The center locations for layers coincide with the location of local minima within the peaks for $\zeta_{0,RMS}$ and are obvious for $\Fr=0.05$ and $\Fr=0.1$. The neighboring local maxima may be used to give a reasonable metric for layer height and indicate the presence of top and bottom sublayers that make up an entire layer. As stratification strength is decreased layer height is observed to increase.
This effect is illustrated in figure~\ref{fig:profile_3} as $\Fr$ is increased from $0.05$ to $0.55$. When $\Fr=0.55$ there is only one large layer of reduced stratification and increased turbulence, and one smaller less-turbulent region of increased stratification that occupies approximately $Z\in [ 0.2, 0.4]$. 

Finally, we note that the instantaneous dissipation rate for energy is increased within the layers.
The instantaneous dissipation rate for horizontal kinetic energy is $Re_f^{-1}\overline{\zeta^2}^\mcl{A}$, and figure~\ref{fig:profile_3}$(c)$ clearly shows that this is increased within the layers.
The dissipation rates for vertical kinetic energy and buoyancy variance are also locally increased within the layers (not shown).
The dynamics leading to the formation of the layers is as yet unknown.

\subsection{Timeseries, equilibration and average energy conversions}\label{sec:timeseries}

\begin{figure}
	\begin{center}
		\begin{tabular}{ccc}
		$(a)$ $Re_{f} = 50,\ \Fr = 0.1414$ & $(b)$ $Re_{f} = 100,\ \Fr = 0.1$ & $(c)$ $Re_{f} = 300,\ \Fr = 0.0577 $ \\ 
		{\rotatebox{90}{\hspace*{.50cm}$\Fr = Re_{f}^{-1/2}$}}
		\includegraphics[width=0.30\linewidth]{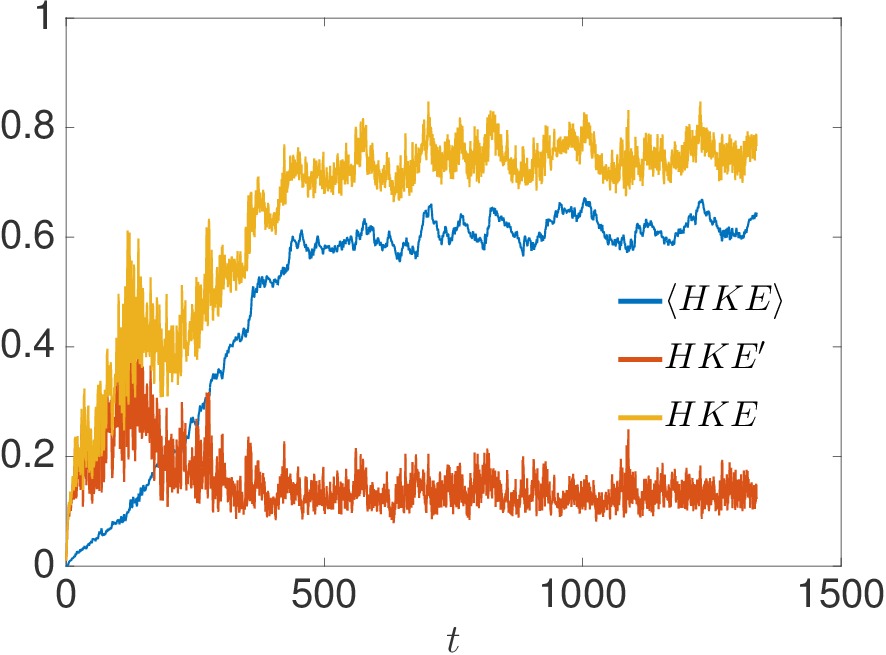} &
		\includegraphics[width=0.30\linewidth]{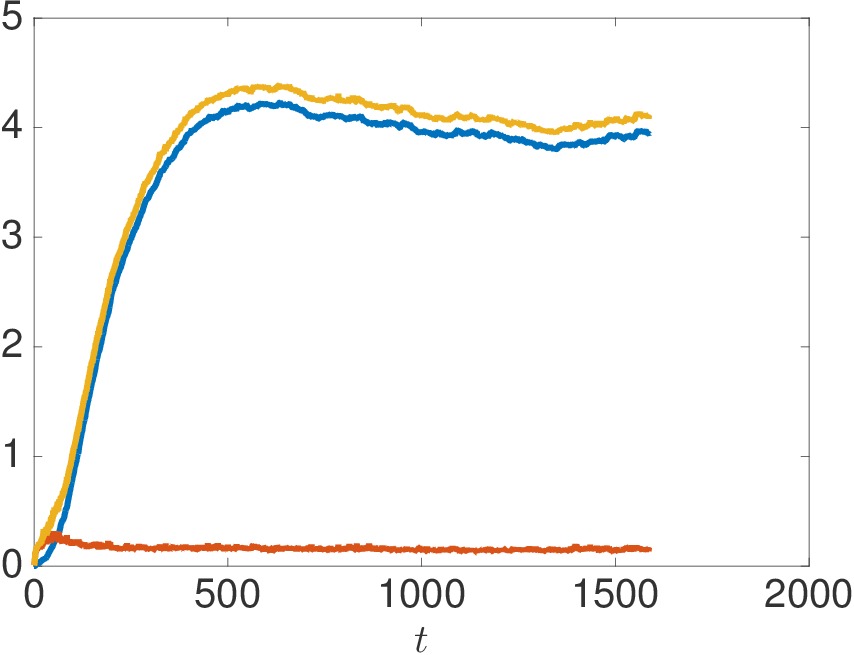} &
		\includegraphics[width=0.30\linewidth]{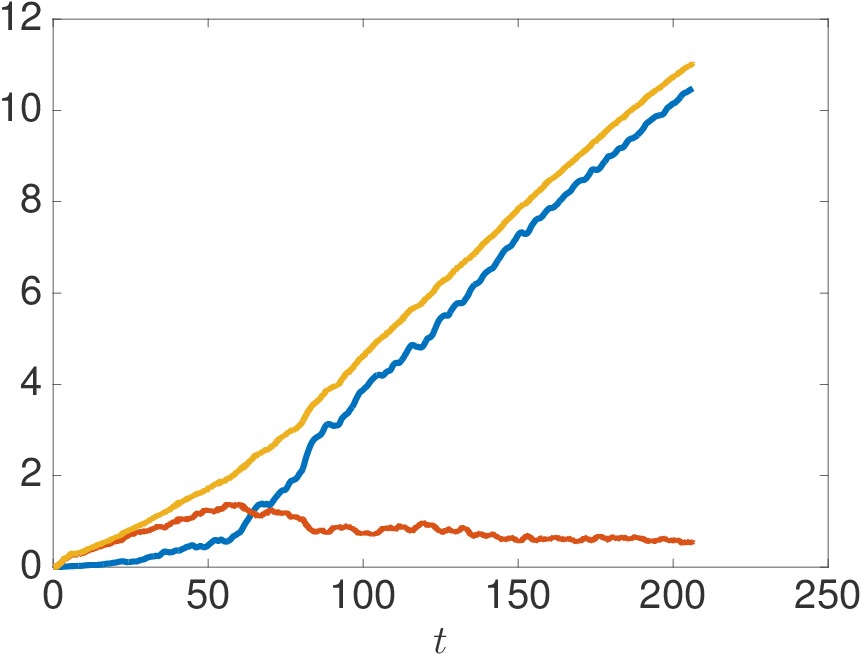} \\
		$(d)$ $Re_{f} = 50,\ \Fr = 18.803$ & $(e)$ $Re_{f} = 100,\ \Fr = 31.623$ & $(f)$ $Re_{f} = 300,\ \Fr = 72.084 $ \\
		{\rotatebox{90}{\hspace*{.70cm}$\Fr = Re_{f}^{3/4}$}}
		\includegraphics[width=0.30\linewidth]{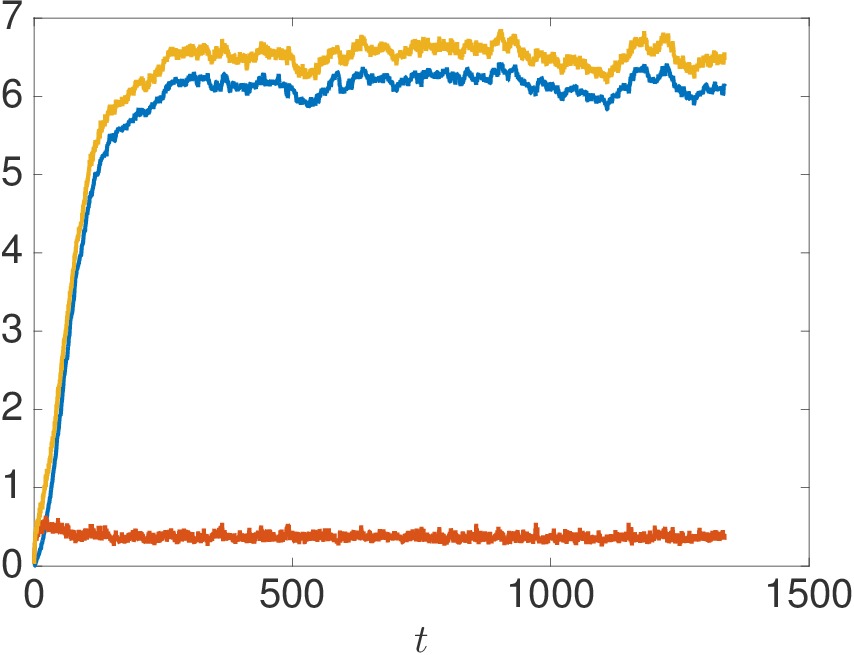} &
		\includegraphics[width=0.30\linewidth]{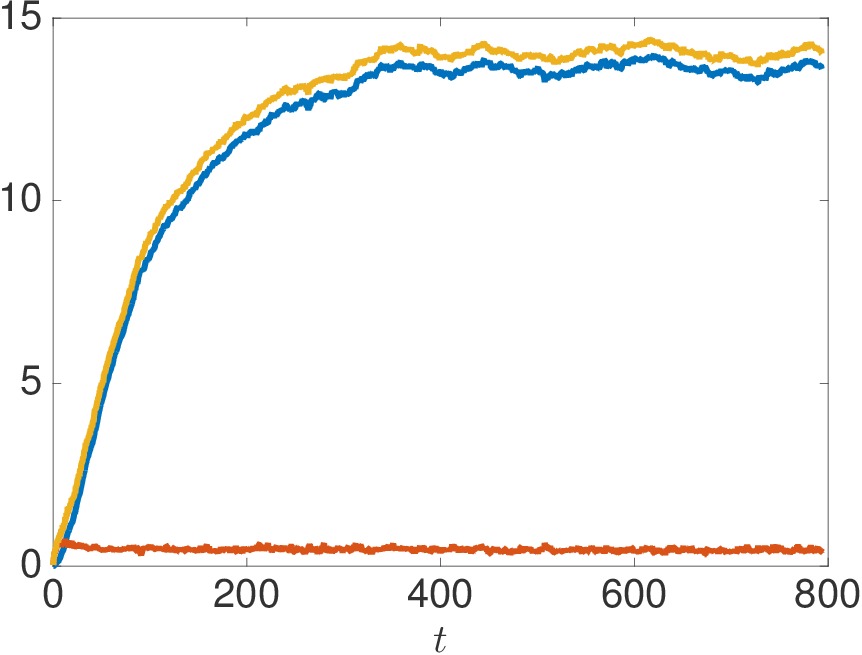} &
		\includegraphics[width=0.30\linewidth]{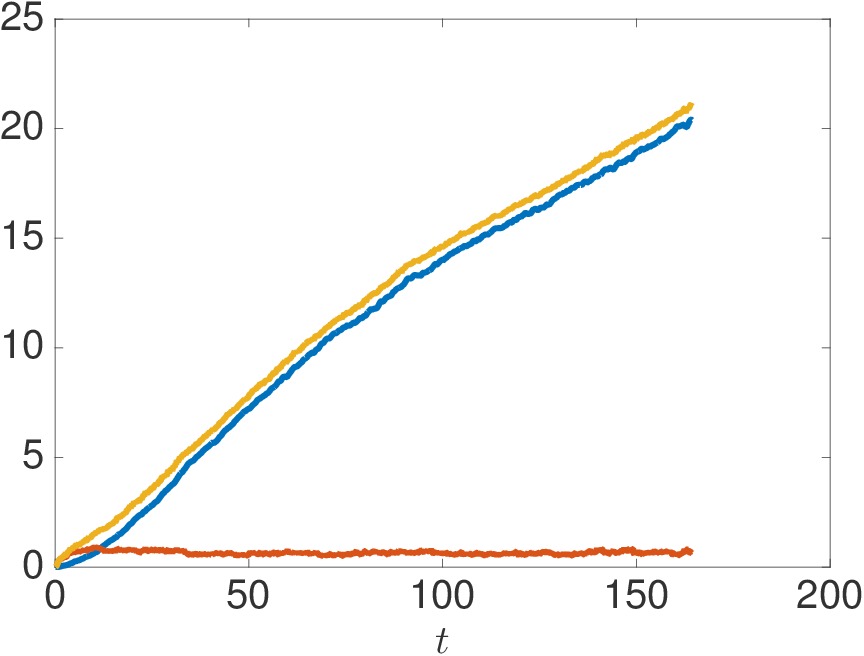} \\
		\end{tabular}
		\caption{Timeseries of volume averaged barotropic, baroclinic and total horizontal kinetic energy at $\Fr = Re_{f}^{-1/2}$ 
		$(a)$-$(c)$ and $\Fr = Re_{f}^{3/4}$ $(d)$-$(f)$. These timeseries correspond to points where an $\times$ sits on
		the dashed lines in figure~\ref{fig:regime_diagram}$(a)$. A notable feature is the saturation of $HKE$ at $Re_{f} = 
		50$ and $Re_{f} = 100$. Computationally expensive simulations at $Re_{f} = 300$ have not equilibrated. The
		barotropic component $\la HKE \ra$ contains nearly all the horizontal kinetic energy after an initial spin-up time.}
		\label{fig:BT_BC_timeseries}
	\end{center}
\end{figure}
We find that total energy is largely dominated by horizontal kinetic energy and this becomes increasingly true as stratification weakens and 
the system approaches purely rotating dynamics. For this reason we focus primarily on the horizontal kinetic energy, hereafter $HKE$.   
Figure~\ref{fig:BT_BC_timeseries} shows timeseries of volume averaged $HKE$ for strong stratification
($\Fr = Re_{f}^{-1/2}$, top row) and weak stratification ($\Fr = Re_{f}^{3/4}$, bottom row) at $Re_{f} = 50,\ 100$, and $300$; the panels correspond to places 
where dashed lines in figure~\ref{fig:regime_diagram} intersect with an $\times$.
Each plot shows the volume averaged barotropic, baroclinic and total horizontal kinetic energy, denoted as $\la HKE\ra$, $HKE'$,
and $HKE$, respectively. 
In every case, the total $HKE$ is dominated by the barotropic part; the only exception in our simulation suite being $Re_{f} = 50$ and $\Fr = 0.0707$, where the energy accumulates in a large vertical scale, but not barotropic (not shown).
At lower Reynolds numbers, $Re_{f}\le100$, the $HKE$ saturates, while the simulations at $Re_{f}=300$ show no indication of saturation, and it is not clear whether it will eventually saturate.

Equation~(\ref{eqn:KEPE}) shows that vortex stretching and vertical buoyancy flux govern the conversion of $VKE$ to $HKE$ and
$PE$ to $VKE$, respectively. Furthermore, conversion of kinetic energy from the baroclinic component $HKE'$ to the barotropic
component $\la HKE\ra$ may be understood by multiplying inviscid equations~(\ref{eqn:BT_BC}$a$) and (\ref{eqn:BT_BC}$b$) by
$-\la\psi_{0}\ra$ and $-\psi'_{0}$ to get

\begin{subeqnarray}
\pd{t}\la HKE \ra  &:=& \pd{t}\left[ \frac{1}{2} \ov{|\nabla_{\perp}\la\psi_{0}\ra|^{2}}^{\mcl{A}}\right]  = 
\ov{\la\psi_{0}\ra \la J[\psi'_{0},\zeta'_{0}]\ra}^{\mcl{A}} \\
\pd{t}HKE'  &:=& \pd{t}\left[ \frac{1}{2} \la \ov{|\nabla_{\perp}\psi'_{0}|^{2}}^{\mcl{A}}\ra\right] =
-\ov{\la\psi_{0}\ra \la J[\psi'_{0},\zeta'_{0}]\ra}^{\mcl{A}} + \la\ov{w'_{0} \pd{Z} \psi'_{0} }^{\mcl{A}} \ra 
\label{eqn:BT_BC_E}
\end{subeqnarray} 

\noindent From the above equations it is clear that vortex stretching occurs only within the baroclinic subspace from which the
two-dimensional barotropic subspace derives its energy. Moreover, flows for which a dynamic equilibrium is obtained have volume
averaged conversion rates that balance dissipation rates. Specifically, by including viscous terms in equations (\ref{eqn:KEPE})
and (\ref{eqn:BT_BC_E}) and assuming steady states, the following expressions for dissipation rates result 

\begin{subeqnarray}
\la HKE_{\mbox{dissip}}\ra&:=&-\ov{\la\psi_{0}\ra \la J[\psi'_{0},\zeta'_{0}]\ra}^{\mcl{A}} =  -\frac{1}{Re_{f}}\ov{\la \zeta_{0}\ra^{2},}^{\mcl{A}}  \\
HKE'_{\mbox{dissip}}&:=&\ov{\la\psi_{0}\ra \la J[\psi'_{0},\zeta'_{0}]\ra}^{\mcl{A}} - \la\ov{w'_{0} \pd{Z} \psi'_{0} }^{\mcl{A}} \ra =
- \frac{1}{Re_{f}}\la \ov{ \zeta_{0}^{\prime 2}}^{\mcl{A}}\ra,  \\
VKE_{\mbox{dissip}}&:=&\la \ov{w_{0}\pd{Z}\psi_{0}}^{\mcl{A}}\ra - \la \ov{w_{0}b_{0}}^{\mcl{A}}\ra - \eps_{f} = - \frac{1}{Re_{f}}\la\ov{|\nabla_{\perp}w_{0} |^{2}}^{\mcl{A}}\ra,\\
PE_{\mbox{dissip}}&:=&\la\ov{w_{0}b_{0}}^{\mcl{A}}\ra  =  -\frac{1}{Pe_{f}} \left\la \frac{\ov{|\nabla_{\perp}b_{0} |^{2}}^{\mcl{A}}}{\pd{Z}\ov{b}_{-1}+\Fr^{-2} S(Z)} \right \ra . 
\label{eqn:BT_BC_diss}
\end{subeqnarray} 

\begin{figure}
	\begin{center}
		\begin{tabular}{cc}
		$(a)\ \la \ov{w_{0}\pd{Z}\psi_{0}}^{\mcl{A}}\ra$ & $(b)\ \ov{\la \psi_{0}\ra \la J[\psi'_{0},\zeta'_{0}]\ra}^{\mcl{A}}$ \\ 
		\includegraphics[width=0.35\linewidth]{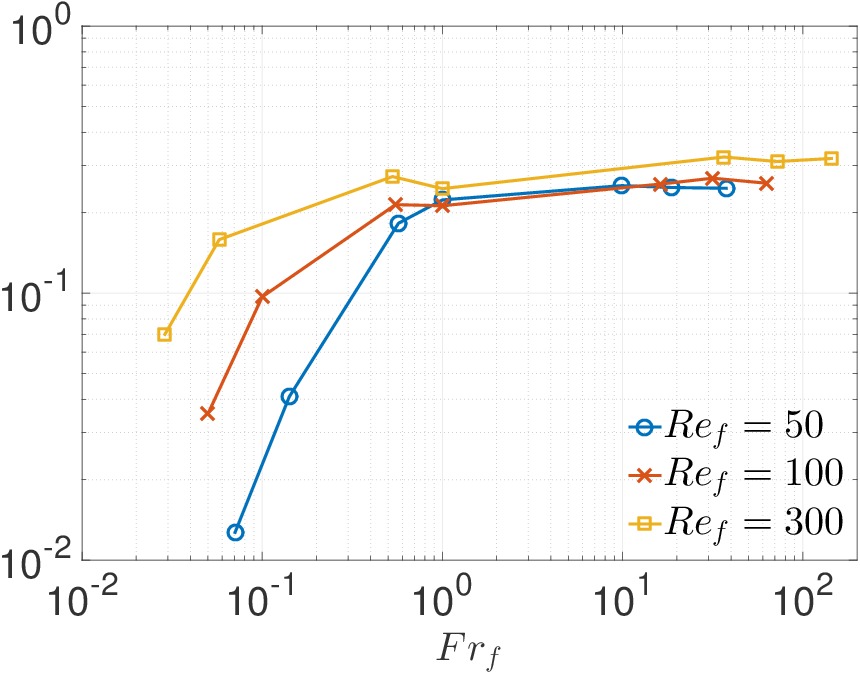} &
		\includegraphics[width=0.35\linewidth]{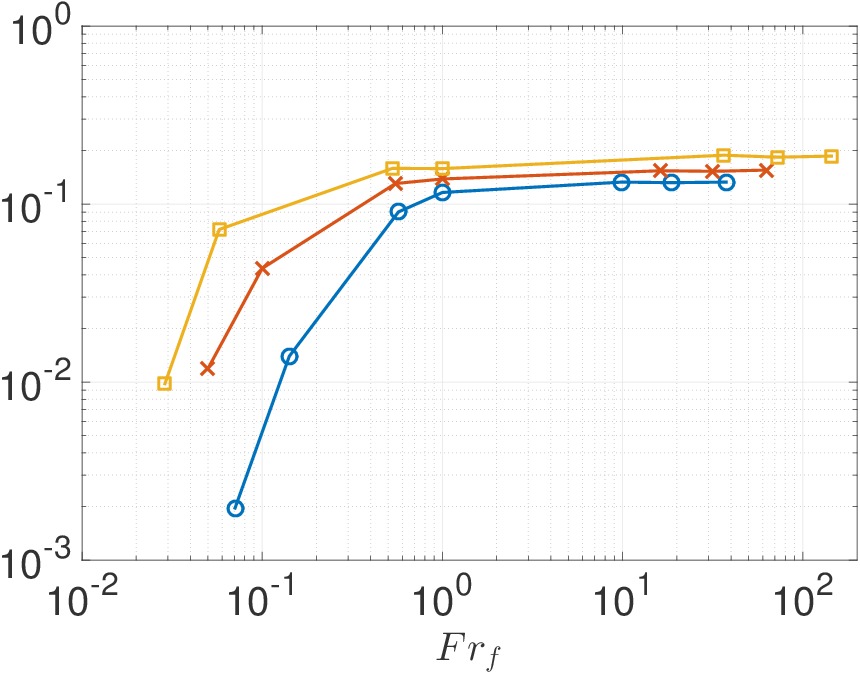} \\
		$(c)\ -\la\ov{ w_{0}b_{0}}^{\mcl{A}}\ra$ &
		$(d)\ \ov{\la \psi_{0}\ra \la J[\psi'_{0},\zeta'_{0}]\ra}^{\mcl{A}} / \la \ov{w_{0}\pd{Z}\psi_{0}}^{\mcl{A}}\ra$ \\
		\includegraphics[width=0.35\linewidth]{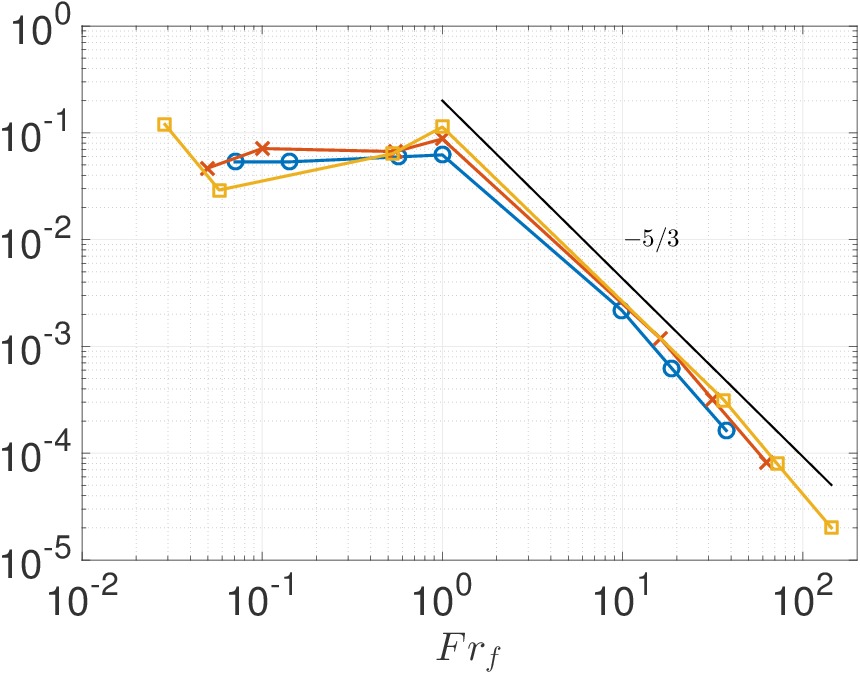} &
		\includegraphics[width=0.35\linewidth]{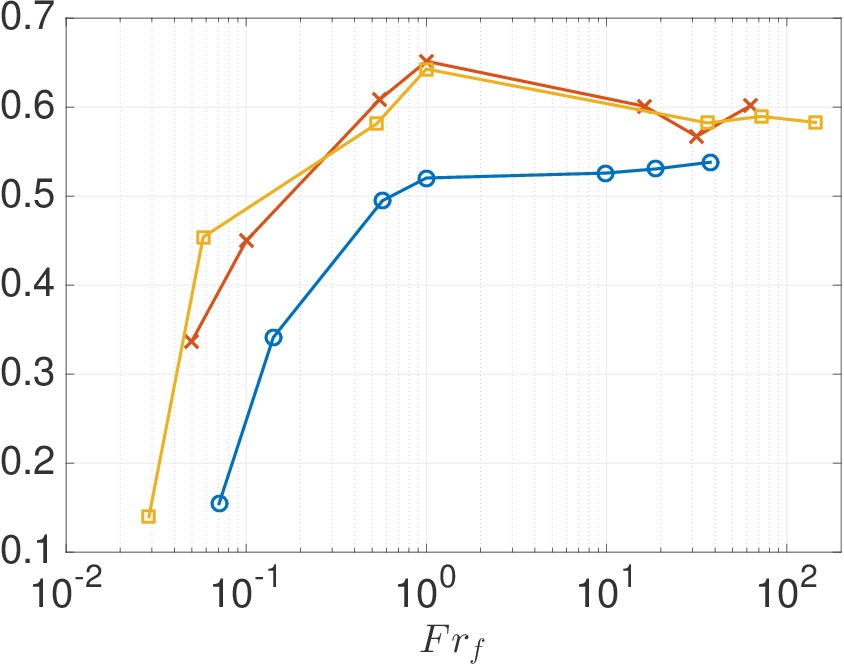} \\
		$(e)\ -\frac{1}{Re_{f}}\la\ov{\zeta_{0}^{\prime 2}}^{\mcl{A}}\ra$ & $(f)\ -\frac{1}{Re_{f}}\la\ov{|\nabla_{\perp}w_{0}|^{2}}^{\mcl{A}}\ra$ \\
		\includegraphics[width=0.35\linewidth]{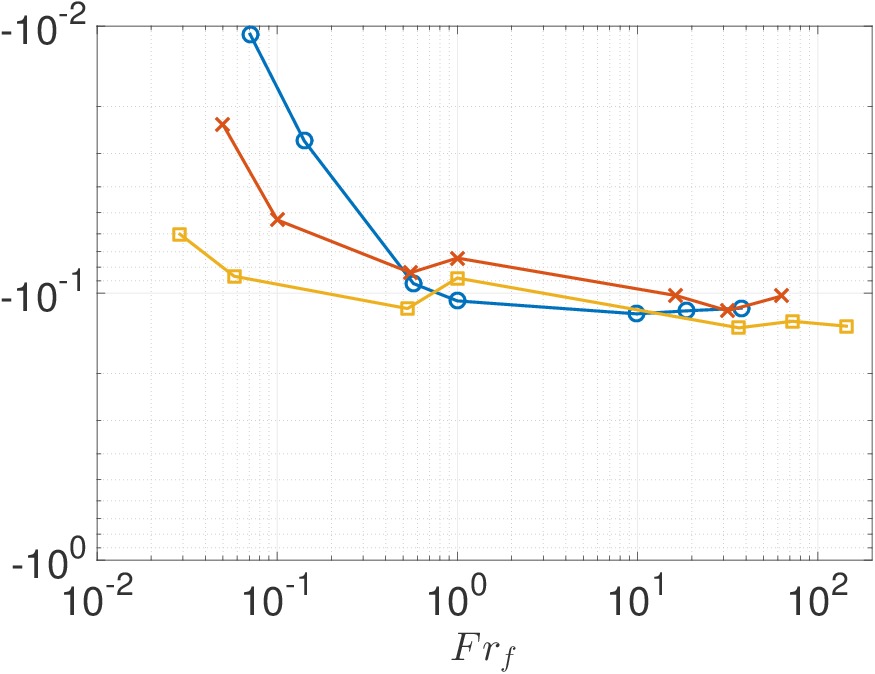} &
		\includegraphics[width=0.35\linewidth]{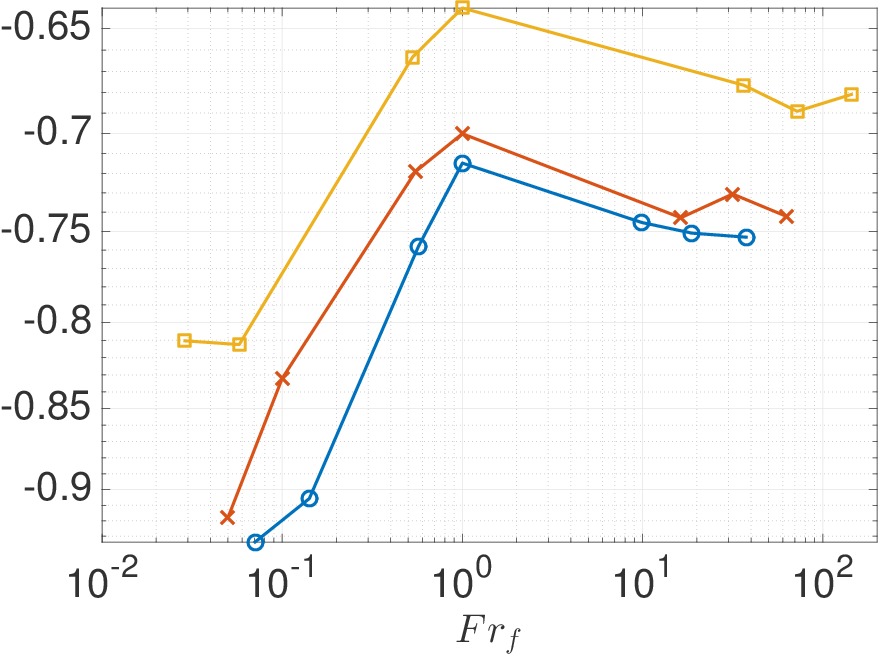} \\
		\end{tabular}
		\caption{Volume and time averaged energy fluxes and dissipation rates for $Re=50,\ 100,\ 300$ for strong ($\Fr < 1$) and weak ($\Fr \geq 1$)
		stratification. 
		Conversion of kinetic energy via $(a)$ vortex stretching (appearing to be most efficient at $\Fr \geq 1$), $(b)$ baroclinic forcing, and $(c)$
		vertical buoyancy flux (showing the decreased role of $PE$ as $\Fr$ increases above $\Fr=1$). Curves in $(d)$ give the ratio of fluxes
		due to baroclinic
		forcing to that due to vortex stretching. Dissipation of $HKE'$ and $VKE$ are given in $(e)$ and $(f)$, respectively. For
		small values of $\Fr$ roughly 90 percent of all energy dissipation is done on $VKE$ while it accounts for about 75 percent
		of energy dissipation at the weakest stratifications. }
		\label{fig:KE_conversion_map}
	\end{center}
\end{figure}

\noindent Summing equations~(\ref{eqn:BT_BC_diss}$a$) and (\ref{eqn:BT_BC_diss}$b$) gives the total dissipation rate of $HKE$, which
matches the total energy conversion by vortex stretching. Summing all dissipation rates in (\ref{eqn:BT_BC_diss}) gives the total energy
dissipation rate, which is precisely the rate $\eps_{f}$ at which
energy is injected. Figure~\ref{fig:KE_conversion_map} shows volume and time averaged energy conversion rates as functions of $\Fr$. These
conversions are those given by equations~(\ref{eqn:KEPE}) and (\ref{eqn:BT_BC_E}). Additionally, for equilibrated flow, as is the case for
simulations with $Re_{f} = 50$ and $Re_{f} =100$, energy conversion rates in figure~\ref{fig:KE_conversion_map} also provide the dissipation
rates given by equation~(\ref{eqn:BT_BC_diss}).
In the following we compare and contrast the ways in which energy is converted from one type to another before being eventually dissipated in the two regimes.

From figure~\ref{fig:KE_conversion_map}$(f)$ it is clear that in both regimes most of the energy input to $VKE$ is dissipated as $VKE$.
A greater percentage of the total energy input is dissipated as $VKE$ in the strong-stratification regime (more than 80\%), but a significant amount is still dissipated as $VKE$ in the weak-stratification regime too (65--75\%).
The vortical mode lacks vertical velocity, and the fact that most of the energy injected to wave modes does not convert to horizontal kinetic or potential energy is an indication of the weakness of the wave-vortex interactions in the rapidly-rotating regime.

Figure~\ref{fig:KE_conversion_map}$(a)$ shows the mean energy conversion rate from $VKE$ to $HKE$ by vortex stretching.
In the weakly-stratified regime the percentage of total energy injection that is converted to $HKE$ remains around 20\%, with a very weak sensitivity to the Reynolds and Froude numbers.
By contrast, as the stratification increases past $\Fr\approx1$ the rate of conversion to $HKE$ drops rapidly, with less conversion for lower Reynolds numbers.
Indeed, of the total input, only approximately 3--4\% is converted to $HKE$ at the smallest Froude number at $Re_f=100$.
This is consistent with known results for the strongly-stratified, rapidly-rotating quasigeostrophic regime where wave modes interact extremely weakly with vortical modes.

We next examine conversion of baroclinic to barotropic $HKE$.
From
equation~(\ref{eqn:BT_BC}) it is clear that baroclinic motions are solely responsible for exciting barotropic motions. In both regimes of weak and strong stratification, we find that the 
conversion of baroclinic to barotropic energy ($F = \ov{\la \psi_{0}\ra \la J[\psi'_{0},\zeta'_{0}]\ra}^{\mcl{A}}$) is roughly statistically steady in time and positive. Time averaged values for the conversion $F$ are summarized in figure~\ref{fig:KE_conversion_map}$(b)$. 
Like the rate of conversion from $VKE$ to $HKE$, the rate of conversion from baroclinic to barotropic $HKE$ is insensitive to $Re_f$ and $\Fr$ in the weakly-stratified regime, and drops sharply with $\Fr$ in the strongly-stratified regime.
Not only does the gross rate of energy injection to the barotropic mode decrease with $\Fr$ in the strongly-stratified regime, the percentage of conversion from $VKE$ to $HKE$ that further converts to barotropic $HKE$ decreases too, as shown by figure~\ref{fig:KE_conversion_map}$(d)$.
For example, at the smallest Froude number and at $Re_f=100$ less than 40\% of the conversion to $HKE$ further converts to barotropic $HKE$.
As mentioned above, the simulation with $Re_{f} = 50$ at the strongest stratification does not exhibit barotropization, which may be due to an insufficient
$\mcl{O}(10^{-3})$ energy flux into the barotropic mode compared to viscous dissipation (see figures~\ref{fig:KE_conversion_map}$(b)$ and \ref{fig:KE_conversion_map}$(e)$).

Clearly, the saturation of the barotropic energy observed at moderate Reynolds numbers is not the result of a shutdown of injection to the barotropic mode.
The fact that the barotropic energy saturates despite a net positive energy injection indicates that there must be a net dissipation to balance the forcing.
None of our simulations use a large-scale dissipation, so the barotropic dissipation must be viscous.
In section \ref{sec:cospectra} we diagnose a small yet robust \emph{direct} cascade of barotropic kinetic energy that carries enough energy to small scale dissipation that the total barotropic energy is able to equilibrate at $Re_{f}\le100$.

Energy injected directly to $VKE$ also converts to potential energy; the mean rate of conversion from $VKE$ to $PE$ is shown in figure~\ref{fig:KE_conversion_map}$(c)$.
This conversion out of $VKE$ displays somewhat opposite behavior to the conversion from $VKE$ to $HKE$: in the strongly-stratified regime the conversion remains flat, insensitive to both Reynolds and Froude numbers, while in the weakly-stratified regime the conversion decreases rapidly as the stratification weakens and with little dependence on Reynolds number.

To summarize, in both regimes energy injected to $VKE$ is primarily dissipated as $VKE$, and there is a net positive conversion to barotropic KE that is, for moderate Reynolds numbers, balanced by dissipation leading to total energy equilibration.
In the strongly-stratified regime the conversion to baroclinic $HKE$ decreases with $\Fr$, as does the rate of conversion to barotropic $HKE$, while the rate of conversion to $PE$ remains moderate and insensitive to $\Fr$.
In the weakly-stratified regime the conversion to baroclinic $HKE$ remains moderate and insensitive to $\Fr$, as does the rate of conversion to barotropic $HKE$, while the rate of conversion to $PE$ decreases rapidly as $\Fr$ increases.

\subsection{Cospectra and scales active in energy conversion}\label{sec:cospectra}

\begin{figure}
	\begin{center}
		\begin{tabular}{ccc}
		$(a)\ (w_{0}, \pd{Z}\psi_{0})$ & $(b)\ (\la \psi_{0} \ra, \la J[\psi'_{0},\zeta'_{0}]\ra)$ & $(c)\ (w_{0},b_{0})$\\ 

		\includegraphics[width=0.32\linewidth]{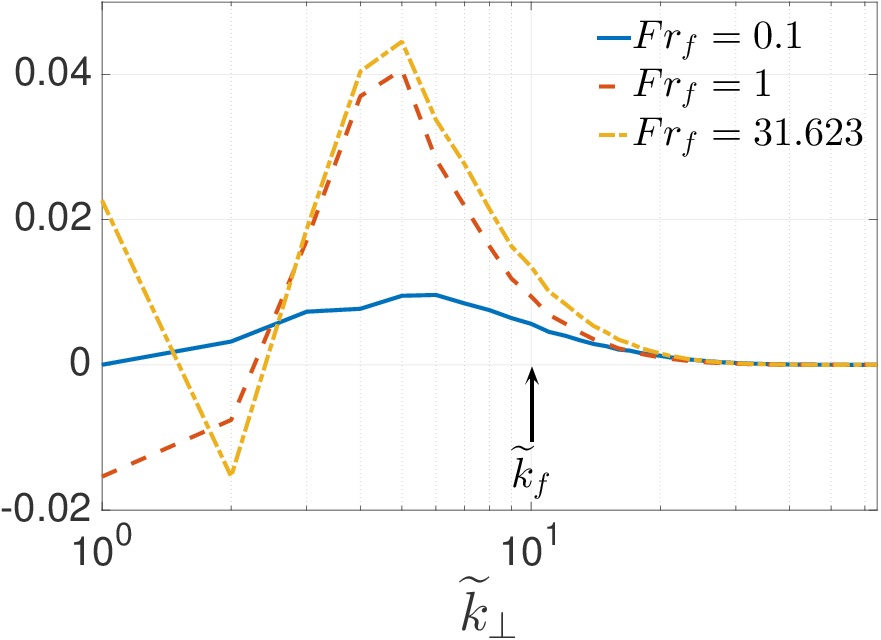} &
		\includegraphics[width=0.32\linewidth]{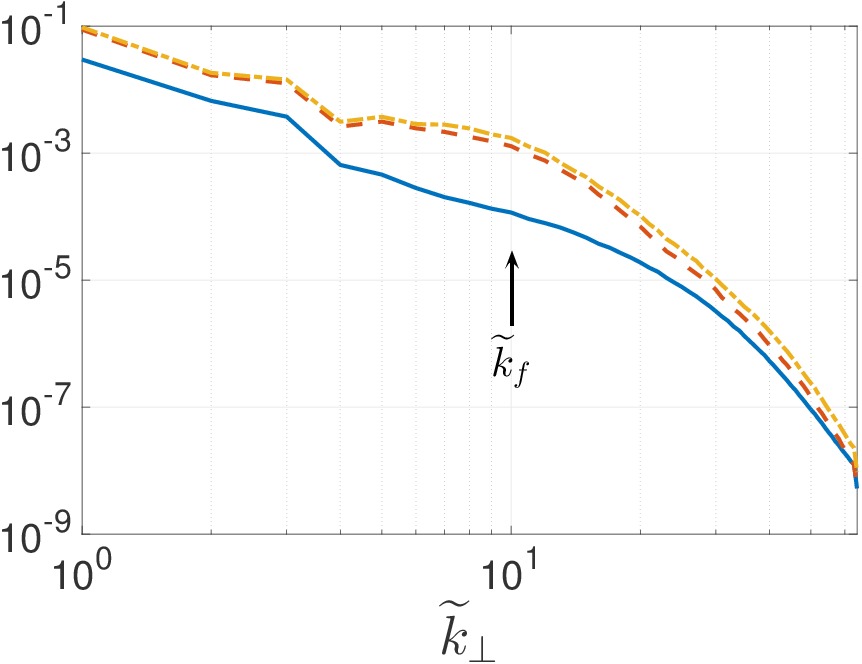} &
		\includegraphics[width=0.32\linewidth]{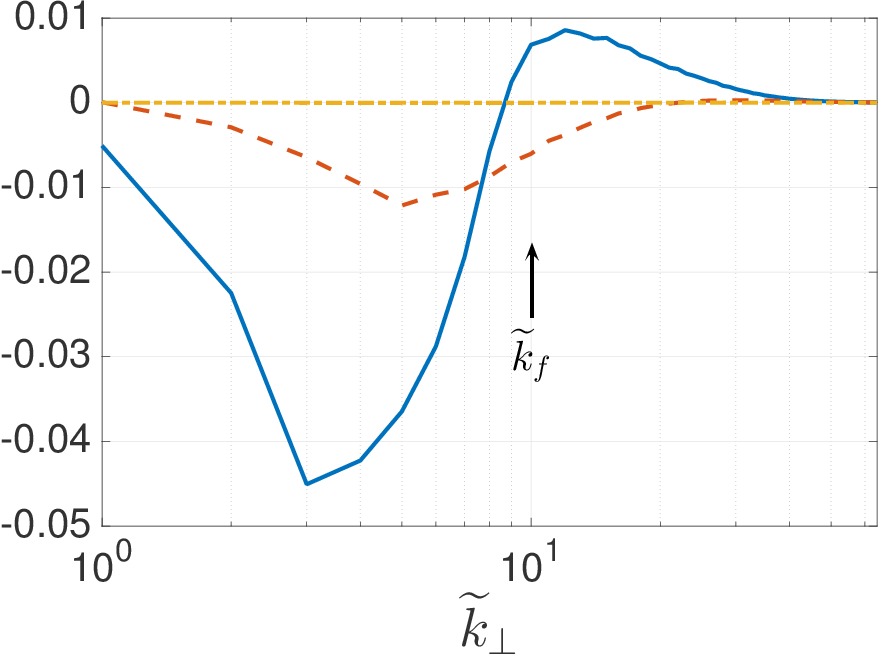} \\
		\end{tabular}
		\caption{Vertically and time averaged horizontal cospectra of energy fluxes at $Re_{f} = 100$ for times proceeding
		energy saturation. Cospectra in $(a)$ give conversions between $HKE$ and $VKE$ by vortex stretching, $(b)$ give
		the barotropization of $HKE$ and indicate a flux of $HKE'$ into the gravest horizontal mode at $\wt{k}_{\perp} = 1$, and
		 $(c)$ give conversions between $VKE$ and $PE$ by vertical buoyancy flux and strongly indicate that these
		 conversion become increasingly weak as $\Fr$ increases beyond unity.}
		\label{fig:cospectra}
	\end{center}
\end{figure}

While illuminating, the discussion in section~\ref{sec:timeseries} is based on global scalars obtained from volume and time averages
and is altogether lacking any spatial information. To improve on this, vertically and time averaged cospectra are computed. These
one-dimensional cospectra are calculated by decomposing horizontal means of point-wise physical space products as a the sum of
Fourier space products, reordering sums over circular rings, binning, and averaging in the $\hat{\mbf{z}}$ direction, i.e.,
  
\begin{equation}
(f,g)(\wt{k}_{\perp})  =\la \sum_{\mathclap{ 0 < | \wt{k}'_{\perp} - \wt{k}_{\perp}| \leq 1}}  \ov{\hat{f}}(k'_{\perp},Z)\hat{g}(k'_{\perp},Z)\ra, 
\quad \wt{k}_{\perp} = \frac{k_{\perp}}{k_{0}} = 1,2,3,\ldots
\label{eqn:cospectra}
\end{equation} 

\noindent where $\wt{k}_{\perp}$ and $\wt{k}'_{\perp}$ are horizontal wavenumbers normalized by the box scale $k_{0}=2\pi/L_{b}$, the bar here denotes
complex conjugation, hats denote horizontal Fourier amplitudes, and angle brackets denote a vertical average. Furthermore, the temporal
mean of cospectra are computed to provide the scales
active in energy conversion on average. Figure~\ref{fig:cospectra} shows cospectra of vortex stretching, barotropization of $HKE$, and
vertical buoyancy flux for simulations with $Re_{f} = 100$ and with $\Fr = Re_{f}^{-1/2},\ 1,\  Re_{f}^{3/4}$. Similar cospectra
are observed for $Re_{f}=50$ and $Re_{f} = 300$. Although simulations with $Re_{f} =300$ have not reached a dynamic equilibrium they too
convey the trends observed for $Re_{f} = 100$ in figure~\ref{fig:cospectra}.

For the strongest stratification, figure~\ref{fig:cospectra}$(a)$ indicates that conversion to $HKE'$ by vortex stretching occurs at all
available scales with a preference for $\wt{k}_{\perp} \approx 5$ (or $L \approx 2L_{f}$), and may hint at
a preferred scale for wave-vortex interactions. The centroid (or the average wavenumber) active for this energy conversion
by vortex stretching is just less than $\wt{k}_{f} = 10$ (or $L \approx L_{f}=1$), however, the efficiency of vortex stretching is best at $L = 2L_{f}$. The
barotropization of $HKE$ in figure~\ref{fig:cospectra}$(b)$ shows that horizontal baroclinic motions act to force barotropic
motions at all scales, however, with a strong preference for the largest available horizontal scale. That this baroclinic forcing is, on average, 
positive definite is consistent with equation~(\ref{eqn:BT_BC_diss}a) and implies that this barotropized energy
is trapped in the barotropic mode until it is viscously dissipated. Figure~\ref{fig:cospectra}$(c)$ shows that the conversion between $PE$ and $VKE$
depends on scale: $VKE$ is converted to $PE$ for $\wt{k}_{\perp} < \wt{k}_{f}$, and
$PE$ is converted back to $VKE$ for $\wt{k}_{\perp} \gtrsim \wt{k}_{f}$ with a net conversion to $PE$ and a peak efficiency at $\wt{k}_{\perp} = 3$
($L \approx 3.3L_{f}$).

When stratification weakens and $\Fr = 1$, there is still a net conversion from $VKE$ to $HKE$ by vortex stretching, but that for stronger stratification stretching converts horizontal kinetic energy back to vertical kinetic energy at the two largest
available scales ($\wt{k}_{\perp} = 1,\ 2$). Vortex
stretching continues to most efficiently convert vertical to horizontal kinetic energy at $\wt{k}_{\perp} = 5$ ($L = 2 L_{f}$), and is nearly four times the 
conversion seen at stronger stratification. Baroclinic 
motions continue to drive barotropic motions in a fashion similar to that at stronger stratification, however, this is done with slightly increased efficiency especially
for $5 < \wt{k}_{\perp} < 20$ (figure~\ref{fig:cospectra}$(b)$). Potential energy becomes weak to the point were the feedback to vertical kinetic
energy for $\wt{k}_{\perp} > \wt{k}_{f}$ is substantially reduced and a preference to convert vertical to potential energy at scales $\wt{k}_{\perp}=5$ ($L=2L_{f}$) is
smaller than that at $\Fr=Re_{f}^{-1/2}$. 


Finally, for the weakest stratification where $\Fr = Re^{3/4}$, conversion from $VKE$ to $HKE$ is very similar to $\Fr=1$, with the exception that
conversion back to vertical kinetic energy only occurs at $\tilde{k}_{\perp} = 2$ rather than both $\tilde{k}_\perp=1$ and $2$. At all other scales vortex stretching acts to move energy from
vertical motions to baroclinic horizontal motions and does so most efficiently near $\tilde{k}_{\perp} = 5$. That the largest scale now plays a role via
vortex stretching in converting vertical to horizontal energy (contrary to what occurs when $\Fr=1$) might be explained by an increased pool of energy made available by the decreased
role of buoyancy (see figure~\ref{fig:KE_conversion_map}$(c)$). Barotropization of horizontal kinetic energy, forced by baroclinic motions, is virtually
identical to $\Fr = 1$ and figure~\ref{fig:cospectra}$(c)$ iterates the insignificance of buoyancy and an approach to purely rotating dynamics. 

\subsection{Energy spectra}\label{sec:hke_spectra}

\begin{figure}
	\begin{center}
		\begin{tabular}{cc}
		$(a)\ (-\psi_{0},\zeta_{0}),\ \Fr=Re_{f}^{-1/2}$ & $(b)\ (-\psi_{0},\zeta_{0}),\ \Fr=Re_{f}^{3/4}$\\ 
		\includegraphics[width=0.48\linewidth]{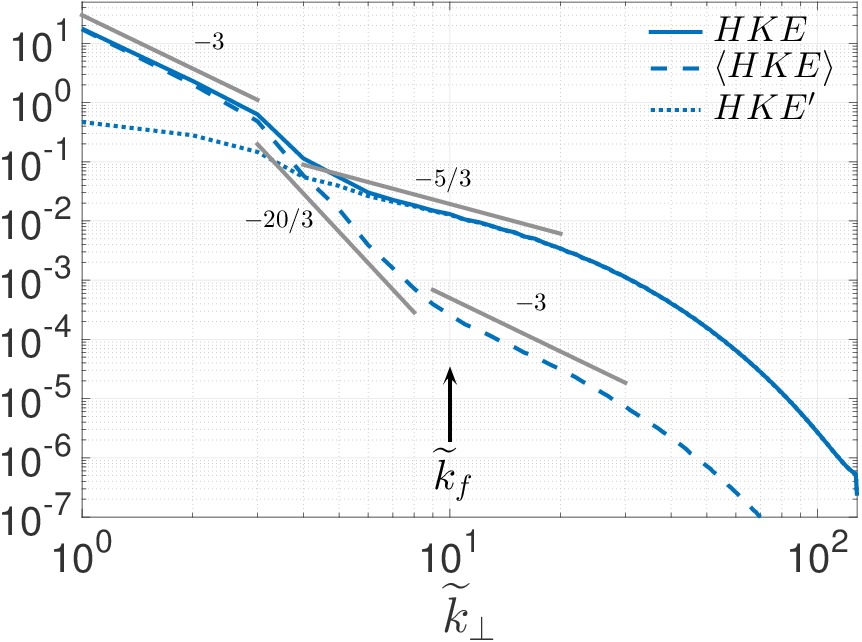} &
		\includegraphics[width=0.48\linewidth]{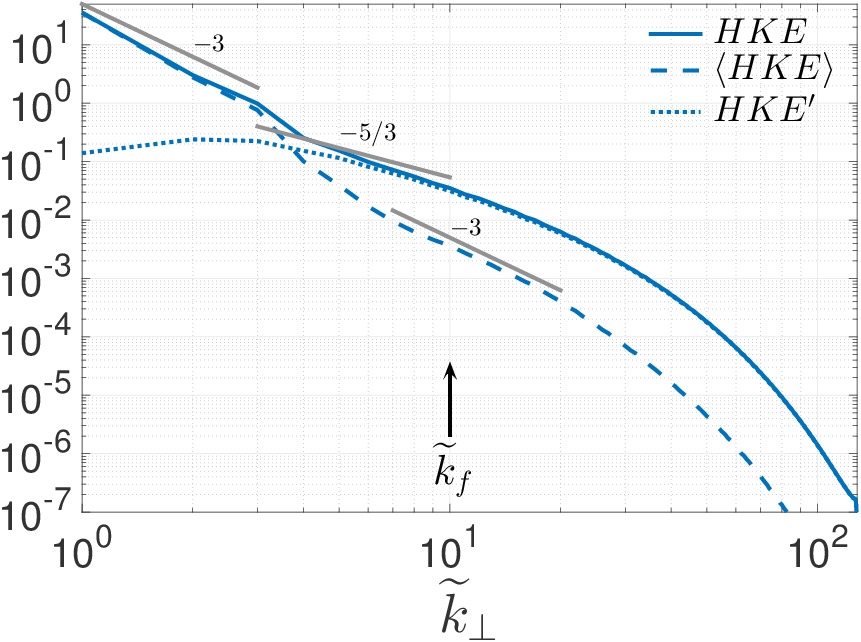} \\
		\end{tabular}
		\caption{Vertically and temporally averaged horizontal energy spectra for $Re_{f} = 300$ with $(a)\ \Fr = 0.0577$ and
		$(b)\ \Fr = 72.084$. Each figure shows the barotropic, $\la HKE\ra = (-\la \psi_{0}\ra,\la \zeta_{0}\ra)$, and baroclinic,
		$HKE' = (-\psi'_{0},\zeta'_{0})$, components of horizontal kinetic energy spectra, $HKE = (-\psi_{0},\zeta_{0})$. A
		$k_{\perp}^{-3}$ energy spectra at small wavenumber is due to energy containing scales in the barotropic subspace
		for both strong and weak stratification. For strong stratification and larger wavenumber a steep $k_{\perp}^{-20/3}$ scaling
		for barotropic energy gives way to a $k_{\perp}^{-3}$ scaling near the dissipation range. For weak stratification the steep
		scaling is short-lived. }
		\label{fig:spectra}
	\end{center}
\end{figure}

Vertically and time averaged horizontal energy spectra for simulations with $Re_{f} = 300$, $\Fr = 0.0577$, and $\Fr = 72.084$
are computed using equation (\ref{eqn:cospectra}) and are given in figure~\ref{fig:spectra}. Similar spectra are observed for remaining values of $\Fr$ and
at lower $Re_{f}$. Both plots give barotropic, $\la HKE\ra = (-\la \psi_{0}\ra,\la \zeta_{0}\ra)$, and baroclinic, $HKE' = (-\psi'_{0},\zeta'_{0})$, components of the total horizontal kinetic energy spectrum, $HKE = (-\psi_{0},\zeta_{0})$. 

For both strong and weak stratification a $\wt{k}_{\perp}^{-3}$ energy spectrum for $\wt{k}_{\perp}\in [1,3]$ is dominated by barotropic energy. For strong stratification the barotropic energy drops off steeply as $\wt{k}_{\perp}^{-20/3}$ for $k_{\perp} \in [3,8]$, and gives way to a $\wt{k}_{\perp}^{-3}$ scaling below the forcing scale. At weak stratification the steep scaling is short-lived and the barotropic spectrum quickly gives way to a $\wt{k}_{\perp}^{-3}$ scaling near the forcing scale. The presence (absence) of the steep drop-off in energy for strong (weak) stratification might be explained,
to some extent, by the weaker (stronger) baroclinic forcing for $\wt{k}_{\perp}\in[3,8]$ (see figure~\ref{fig:cospectra}$(b)$), indeed, the shape of the
forcing cospectrum decreases (sustains) in this range. In turn, the difference in behavior of baroclinic forcing might be explained by flow morphology.
At strong stratification horizontal layers appear and are associated with increased viscous effects that may disrupt collinearity of baroclinic advection of the baroclinic vorticity with the barotropic streamfunction (figure~\ref{fig:cospectra}$(b)$). When layers are absent
at weaker stratification so are associated regions of increased viscous effects and the result is an increased efficiency of baroclinic forcing
(figure~\ref{fig:cospectra}$(b)$). 

For strong stratification, as $\wt{k}_{\perp}$ increases and barotropic energy becomes subdominant, the baroclinic energy spectra scales as $\wt{k}_{\perp}^{-5/3}$ for $\wt{k}_{\perp} \in [4,\approx 20]$. When stratification is weaker this scaling range appears to narrow, which may be explained by
increased vortex stretching which acts to force baroclinic energy most efficiency in the range $\wt{k}_{\perp}\in [4,5]$.

\subsection{Barotropization and inverse cascade}\label{barotropization}

\begin{figure}
	\begin{center}
		\begin{tabular}{c}
		$T_{\wt{k}_{\perp}\wt{p}_{\perp}},\  T_{\wt{k}_{\perp}}$ \\ 
		\includegraphics[width=0.45\linewidth]{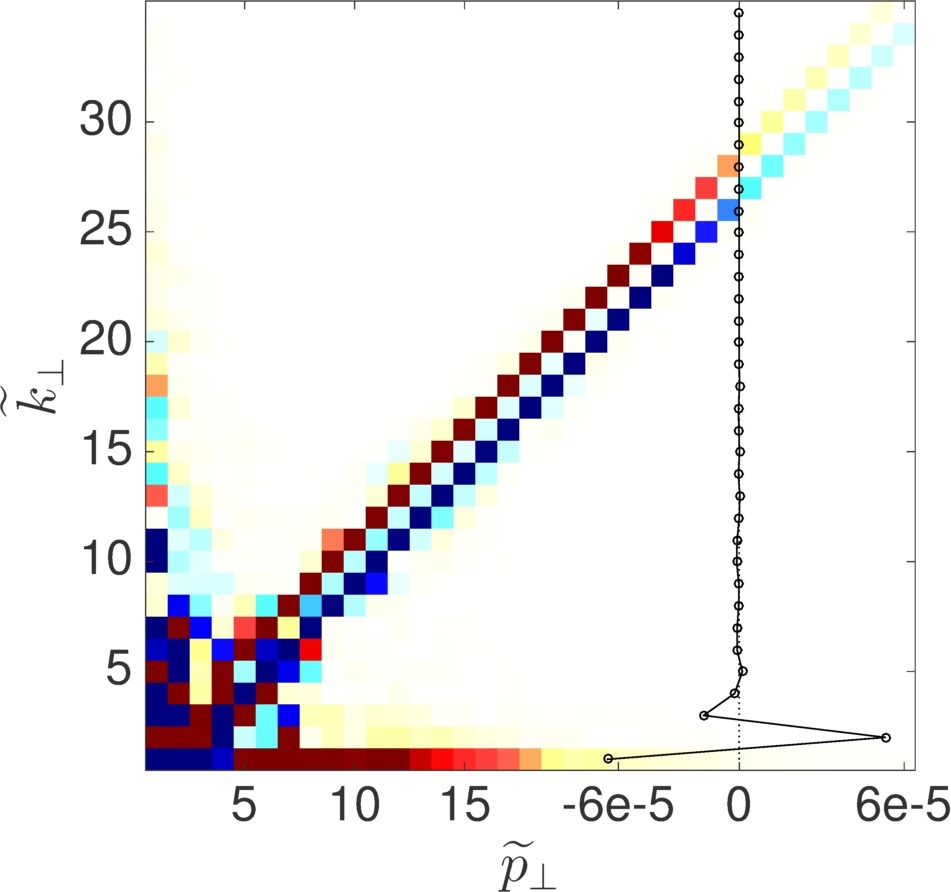} \\
		\end{tabular}
		\caption{Energy transfer map showing how barotropic triad interactions move energy within the barotropic subspace
		for equilibrated dynamics where $Re_{f} = 100$ and $\Fr = 0.1$. The vertical profile (on the right) is the result of summing
		the transfer map $T_{k_{\perp}p_{\perp}}$ over $p_{\perp}$ to get $T_{k_{\perp}}$. Note the scale for $T_{k_{\perp}}$ is 
		$\mcl{O}(10^{-5})$, an indication that energy transfer via triad interaction are weak relative to baroclinic forcing. Similar
		results are seen for weak stratification. Red (blue) shading indicates that energy is transferred into (out of) wavenumber $\wt{k}_{\perp}$
		through interactions with wavenumber $\wt{p}_{\perp}$. }
		\label{fig:BT_transfer}
	\end{center}
\end{figure}

It is interesting to consider the barotropic dynamics since these motions are governed by the two-dimensional vorticity equation (\ref{eqn:BT_BC}a).
If two-dimensional dissipative flow is forced at scales well separated from frictional effects acting on energy and enstrophy then an upscale energy range and a downscale enstrophy range form where, in the limit of vanishing viscosity the downscale transfer of energy through the enstrophy range is expected to vanish. In our simulations, of the energy converted to baroclinic $HKE$ by vortex stretching a fraction (which depends on $\Fr$) acts to force the barotropic vorticity equation (see figure~\ref{fig:KE_conversion_map}$(d)$). Figure~\ref{fig:cospectra}$(b)$ illustrates that baroclinic
motions establish a natural injection of energy directly into the gravest barotropic mode, so that the accumulation of energy at large scales in the barotropic mode does not result primarily from a two-dimensional inverse-cascade process.
Dissipation in the barotropic subspace, therefore, occurs through a nonzero forward energy cascade. Figure~\ref{fig:BT_transfer} gives a detailed map of the transfer of energy between barotropic Fourier modes performed
by barotropic triad interactions for equilibrated flow at $Re_{f} = 100$ \citep{aR14}. The near-diagonal elements of this map at large wavenumber show a local forward transfer of energy to small scales coexisting with a non-local inverse cascade at larger scales.
The accumulation of energy at large horizontal scales in the barotropic mode in these rapidly rotating flows is primarily the result of three-dimensional baroclinic motions interacting to directly induce large-scale and vertically-invariant structures; it is not primarily a result of baroclinic injection to an intermediate scale, followed by a purely-barotropic inverse cascade to larger scales.

%
%
%

\section{Conclusions}\label{sec:conclusions}

We have presented an investigation into stably stratified and rapidly rotating turbulence using the asymptotically reduced NH-QG equation set valid for $Ro\ll 1$ describing geostrophically balanced flow. Such a regime is relevant to abyssal oceans (where observations indicate the presence of weak stratification) as well as planetary and stellar interiors (in regions where stratification transitions from unstable to stable). Within this parameter regime the Proudman-Taylor constraint is relaxed/broken by allowing anisotropic dynamics with vertical scales $\mcl{O}(Ro^{-1})$ larger than horizontal scales. In this setting slow inertia-gravity waves with order-one frequencies are retained and not filtered, moreover, timescales for nonlinear eddy dynamics and anisotropic inertia-gravity waves are not asymptotically separated (see \S~\ref{sec:TP_relax}). Numerical simulations with wave-eddy interactions are performed where motions are induced by a stochastic injection of vertical kinetic energy; doing so only provides wave-energy and any emergence of vortical mode energy must originate from wave-eddy interactions (see \S~\ref{sec:num_meth}). 

Our results reveal two regimes corresponding to strong ($\Fr<1$) and weak ($\Fr\geq1$) stratification. These regimes are primarily distinguished by the presence at strong stratification of thin horizontal turbulent layers in which energy transfer and dissipation are most active. As $\Fr$ increases up to unity, layer thickness also increases until the layers occupy the entire vertical extent of the domain. We note such layer formation is not observed for classical QG dynamics for which inertia-gravity waves are entirely absent. Evidence of layering has been previously observed in experiments of decaying purely stratified turbulence \citep{BC2000} and 
numerical studies of decaying rotating-stratified turbulence \citep{Cambon2001}, but not in previous studies of rapidly-rotating, strongly-stratified, forced-dissipative turbulence. Unlike the `pancake' structures that form in stratified turbulence \citep{KH12}, the layers here are localized and long-lived. Also, vertical shear of the horizontal velocity $\pd{z}u_\perp$ is absent from the reduced equations governing the dynamics, so layer formation cannot be associated with shear instabilities like Kelvin-Helmholtz or symmetric instability. Unlike the `staircase' layering in doubly-diffusive convection \citep{STGBR11} the layers consist of thin regions of \emph{reduced} stratification. We conjecture that their existence is related to our use of vertical velocity forcing, in the sense that other kinds of forcing may disrupt the dynamics leading to layer formation. Here, layer formation at $\Fr<1$ is associated with mixing by vertical buoyancy flux and energy conversion by vortex stretching (evident in vertical profiles in figure~\ref{fig:profile_3}). Additionally, vertical profiles of stratification and RMS vertical vorticity quantify layer location and thickness. 

In addition to the presence or absence of layers, the regimes are distinguished by energetics.
In the strongly-stratified regime only a small percentage of the energy injection rate to vertical kinetic energy is converted to horizontal kinetic energy, and a modest amount is converted to potential energy.
In the weakly-stratified regime only a small percentage of the energy injection rate to vertical kinetic energy is converted to potential energy, and a modest amount is converted to horizontal kinetic energy.

Both regimes are characterized by the emergence of a large-scale barotropic dipole (see figure~\ref{fig:regime_diagram}). When the Reynolds number is not too large ($Re_{f} \leq 100$, or $Re_{c} \leq 2000$) system energy reaches a statistically steady state, evidence that geostrophically balanced flow is capable of establishing a direct route to dissipation. The process leading to energy saturation is attributed to a downscale transfer of kinetic energy within the barotropic mode, which balances the injection of barotropic energy by baroclinic motions. Another distinct trait of the flows studied here is that three-dimensional baroclinic motions interact in such a way as to inject energy into the largest barotropic scales; therefore, the accumulation of energy at the largest scales in the barotropic mode is not the result of an upscale transfer within the barotropic mode.\\

The authors would like to thank Peter Bartello and  Mary-Louise Timmermans for helpful discussions. This work was supported by the National
Science Foundation under grants EAR CSEDI \#1067944  and DMS \#1317666 (KJ), DMS \#1444503 (DN), and in part by NSF-OCE 1245944 (JBW).
This work utilized the Janus supercomputer, 
which is supported by the National Science Foundation (award number CNS-0821794) and the University of Colorado Boulder. The Janus supercomputer is 
a joint effort of the University of Colorado Boulder, the University of Colorado Denver and the National Center for Atmospheric Research.

\bibliographystyle{jfm}
\bibliography{JFM_NJ_2015.bib}

\end{document}